\documentclass{article}
\usepackage[letterpaper, left=1in, right=1in, top=1in, bottom=1in]{geometry}
\usepackage{authblk}
\usepackage{cite}
\usepackage{amsmath,amssymb}
\usepackage{amsfonts}
\usepackage{bm}
\usepackage{dirtytalk}
\usepackage{hyperref}
\usepackage{xcolor}
\usepackage{graphicx}
\usepackage{tikz}
\usetikzlibrary{positioning, fit, arrows, arrows.meta, shapes.geometric}
\tikzstyle{cnn} = [rectangle, minimum width=2.cm, minimum height=0.4cm, text centered, draw=black]
\tikzstyle{cnnr} = [rectangle, minimum width=1.75cm, minimum height=0.4cm, text centered, draw=black, rotate=90]
\tikzstyle{circ} = [circle, minimum width=.0em, minimum height=0.55cm, text centered, draw=black]
\tikzstyle{arrow} = [-{Stealth[scale=.75]}]
\tikzstyle{revarrow} = [{Stealth[scale=.75]}-]
\definecolor{azul}{RGB}{187, 233, 255}

\title{Super-Resolution Emulation of Large Cosmological Fields with a 3D Conditional Diffusion Model}

\author[1]{Adam Rouhiainen\thanks{Electronic address: rouhiainen@wisc.edu}}
\author[2]{Michael Gira}
\author[1]{Moritz M{\"u}nchmeyer}
\author[2]{Kangwook Lee}
\author[1]{Gary Shiu}

\affil[1]{Department of Physics, University of Wisconsin-Madison, Madison, WI 53706, USA}
\affil[2]{Department of Electrical and Computer Engineering, University of Wisconsin-Madison, Madison, WI 53706, USA}

\date{\small November 9, 2023}

\begin{document}

\maketitle

\begin{abstract}
    High-resolution (HR) simulations in cosmology, in particular when including baryons, can take millions of CPU hours. On the other hand, low-resolution (LR) dark matter simulations of the same cosmological volume use minimal computing resources. We develop a denoising diffusion super-resolution emulator for large cosmological simulation volumes. Our approach is based on the image-to-image Palette diffusion model, which we modify to 3 dimensions. Our super-resolution emulator is trained to perform outpainting, and can thus upgrade very large cosmological volumes from LR to HR using an iterative outpainting procedure. As an application, we generate a simulation box with 8 times the volume of the Illustris TNG300 training data, constructed with over 9000 outpainting iterations, and quantify its accuracy using various summary statistics.
\end{abstract}

\section{Introduction}

Cosmological simulations of dark matter and baryonic matter are crucial for modern cosmology. They are required for theoretical studies, statistical method development, and parameter inference on real data. Unfortunately, high-resolution (HR) simulations with baryonic matter, even on moderate cosmological volumes, can require millions of CPU hours. On the other hand, low-resolution (LR) simulations of dark matter can easily be generated using much smaller computational resources. Our approach is to train a 3-dimensional diffusion model to upgrade LR simulations to super-resolution (SR) simulations, which emulate the HR simulations at our targeted resolution. Because the small scales of the HR simulation are not contained in the LR simulation, an SR emulator should be probabilistic, able to generate many SR simulations consistent with the same LR simulation. We thus require a stochastic generative model, which learns probabilistic small-scale physics from HR training data. In this work, we choose a denoising diffusion probabilistic model~\cite{sohl2015, ho2020denoising, palette}, because they are currently the best performing models for image generation, generally outperforming GANs on various image tasks ~\cite{dhariwal2021diffusion}, and are much more expressive than normalizing flows in higher dimensions~\cite{Papamakarios2021,kong2020expressive}.

This work develops a stochastic treatment of super-resolution emulation combined with a conditional outpainting procedure that can make large volumes without boundary effects. Our outpainting method generates new SR volumes conditioned both on the LR simulation and on previously generated neighboring SR volumes, inspired by auto-regressive models such as PixelCNN~\cite{van2016conditional}. By repeatedly outpainting smaller, local volumes many times over, we are able to generate simulation volumes larger than the entire HR training data volume. This is possible due to the locality of the underlying physics of structure formation, where on large scales, LR and HR simulations with the same initial conditions give the same final result. Therefore, we are able to emulate features with length scales larger than the outpainting window so long as such features are in the LR conditional. Details of our approach and its alternatives are discussed in Sec.~\ref{sec:iterative_outpainting}.

There have been various previous approaches to super-resolution of cosmological data. In~\cite{kodiramanah2020}, the authors trained a Wasserstein GAN as a super-resolution emulator at field level, in a similar setup to our work. However, their approach did not probabilistically model the SR conditioned on the LR simulation. Instead, they generate a deterministic SR solution, conditioned on the HR initial conditions field. This is an interesting and efficient approach but does not allow to explore the probabilistic space of consistent SR simulations given an LR simulation. In a different series of works~\cite{li2021_1, li2021_2, li2023}, a GAN is used to generate SR simulations stochastically at particle level, increasing the particle count in the simulation boxes by several orders of magnitude. This is a powerful alternative approach that is closer to N-body simulations, which naturally work with particles. We believe that both particle and field-level approaches can be useful. Cosmological survey analysis is usually done on a regular grid (for example, to take FFTs), so that particle positions are not necessarily required. On the other hand, the Lagrangian approach has the advantage that it can adaptively represent a physical field with a large dynamic range.
Recently,~\cite{schanz2023stochastic} used a diffusion model to perform super-resolution on 2-dimensional dark matter fields. Their work is technically most similar to ours, with their approach including a Fourier filter on the large-scale structure data to boost the importance of learning the nonlinear scales. Our work is volumetric and also includes an auto-regressive procedure to generate large volumes by conditioning sub-volumes on their neighbors.

Diffusion models have been used in astrophysics for other purposes than super-resolution emulation. They have been used to generate astrophysical fields projected to 2 dimensions~\cite{Mudur:2022gfq, Zhao:2023giv}, and reconstruct strong gravitational lensing images~\cite{Karchev:2022ycy}. Further, a 3-dimensional diffusion model was recently used to infer initial conditions from present day dark matter simulations, using data meshed onto $128^3$ voxels~\cite{10.1093/mnrasl/slad152}.

This paper is organized as follows. We briefly review the theory behind denoising diffusion models in Sec.~\ref{sec:diffusion_theory}. In Sec.~\ref{sec:data}, we describe the HR and LR training data, and LR test set data, used in this project. In Sec.~\ref{sec:diffusion_model}, we describe the diffusion model architecture with our iterative outpainting method. In Sec.~\ref{sec:results}, we show our results for a generated field larger than the entire training data volume. In Sec.~\ref{sec:results_variety}, we use the stochastic nature of diffusion models to generate a variety of SR fields conditional on a single LR field. In Appendix~\ref{sec:BAO_response}, we demonstrate SR emulation of baryon acoustic oscillations, which have length scales larger than our model's outpainting window. In Appendix~\ref{sec:variable_cosmos}, we test the robustness of the diffusion mode to slightly out-of-distribution data by varying the cosmological parameters of the LR field, and generating new SR fields.

\section{Diffusion model theoretical background}
\label{sec:diffusion_theory}

The image-to-image denoising diffusion probabilistic model is a stochastic generative model that samples $\bm{y}$ from $P{(\bm{y}|\bm{x})}$, where the image $\bm{y}$ is conditional on a (usually low-resolution) image $\bm{x}$. Applications of image-to-image diffusion include colorization, denoising, JPEG resoration, inpainting, outpainting, and other types image reconstruction and super-resolution. Here we briefly explain the theory behind this conditional diffusion model, with more details available in~\cite{ho2020denoising}. The equations here are valid in any number of spacial dimensions.

The diffusion model has a forward process for training, and a reverse process for generation. In the forward diffusion step, a field $\bm{y}_t$ gets added noise from a field $\bm{y}_{t-1}$ as
\begin{equation}
    q(\bm{y}_t|\bm{y}_{t-1})=\mathcal{N}{(\bm{y}_t|\sqrt{1-\beta_t}\ \bm{y}_{t-1},\beta_t\mathbb{I})},
\end{equation}
where $\beta_t$ control the amount of noise reduction between steps. By iterating this forward diffusion process, the field $\bm{y}_t$ at some step $t$ can be written in terms of the original field $\bm{y}_0$ as
\begin{align}
    q(\bm{y}_t|\bm{y}_0)&=\prod_{i=1}^tq{(\bm{y}_i|\bm{y}_{i-1})}\\
    &=\mathcal{N}{(\bm{y}_t|\sqrt{\gamma_t}\ \bm{y}_0,(1-\gamma_t)\mathbb{I})}
\end{align}
where $\gamma_t=\prod_{i=1}^t(1-\beta_i)$. A total of $T$ steps are taken. The $\gamma_t$ are chosen as model hyperparameters in such a way that by the end of the diffusion chain, $\sqrt{\gamma_T}\ \bm{y}_0$ is small relative to $\bm{y}_T$, and thus the noisy data has lost almost all resemblance to $\bm{y}_T$. % The reverse diffusion step, given $\bm{y}_0$, is

The purpose of diffusion models is to generate fields in the distribution $\bm{y}_0$ without knowing the true value of $\bm{y}_0$, and so we need an equation for $\bm{y}_{t-1}$ in terms of $\bm{y}_t$ without $\bm{y}_0$. To achieve this, we fit a function $f_{\bm{\theta}}{(\bm{x},\bm{y}_t,\gamma_t)}$ to the noise $\bm{\epsilon}\sim\mathcal{N}(\bm{0},\mathbb{I})$ by minimizing the loss function
\begin{equation}
    \label{eq:loss}
    \mathbb{E}_{(\bm{x},\bm{y})}\mathbb{E}_{\bm{\epsilon},\gamma}\left\Vert f_{\bm{\theta}}(\bm{x},\sqrt{\gamma}\ \bm{y}_0+\sqrt{1-\gamma}\ \bm{\epsilon},\gamma)-\bm{\epsilon}\right\Vert_2^2
\end{equation}
where $\left\Vert...\right\Vert_2^2$ is the squared $L_2$ norm. In practice, the function $f_{\bm{\theta}}{(\bm{x},\bm{y}_t,\gamma_t)}$ is a neural network with parameters $\bm{\theta}$, and the details of our network are described in Sec.~\ref{sec:network_architecture}.

There are multiple ways to solve for the reverse diffusion step~\cite{song2022denoising,lu2022dpmsolver}, and we have a brief discussion of these various approaches in Sec.~\ref{sec:conclusion}. In this work, we use the results of \cite{ho2020denoising,palette} to write the reverse diffusion step as
\begin{equation}
    \bm{y}_{t-1}=\frac{1}{\sqrt{1-\beta_t}}\left(\bm{y}_t-\frac{\beta_t}{\sqrt{1-\gamma_t}}f_{\bm{\theta}}{(\bm{x},\bm{y}_t,\gamma_t)}\right)+\sqrt{\beta_t}\ \bm{\epsilon}_t.
\end{equation}
This reverse diffusion step is repeated to obtain a sample $\bm{y}_0$ from $\bm{y}_T$. Because the starting field $\bm{y}_T$ is a random Gaussian field, we can obtain a variety of samples $\bm{y}_0$ without changing the network parameters $\bm{\theta}$.

\section{Data}
\label{sec:data}

Here we describe the HR TNG300 training data and its accompanying LR conditional training data. We also describe our the LR conditional test data, which has different initial conditions (phases), and a larger volume for Sec.~\ref{sec:results}.

\subsection{High-resolution data from TNG300}

The HR data to learn the baryonic physics of the large-scale structure comes from IllustrisTNG~\cite{tng2017_1,tng2017_2, tng2017_3,tng2017_4,tng2017_5}, a set of three gravo-magnetohydrodynamical simulations run on \texttt{Arepo}~\cite{Springel_2010}. We picked TNG300 because it covers a large enough volume to include both linear and nonlinear physics, and to demonstrate that a single HR simulation can be enough to train a model that generates larger volumes. The TNG300 run  simulated $2500^3$ baryon particles and $2500^3$ dark matter particles in a $205\ \text{Mpc}/h\approx300\ \text{Mpc}$ length cube, with cosmological parameters $\Omega_\Lambda=0.6911$, $\Omega_\text{M}=0.3089$, $\sigma_8=0.8159$, $n_\text{s}=0.9667$, and $h=0.6774$. The simulation has an initial field at redshift of $z=127$ generated with \texttt{NGenIC}, using the Zel'dovich approximation. TNG300 computed over $10$ million time steps down to $z=0$, and we use the $z=0.01$ snapshot. We use the cloud-in-cell mass assignment scheme to place the TNG300 baryons onto a $264^3$~px cubic mesh. We will comment on increasing this resolution in Sec. \ref{sec:conclusion}. Our target field in the present work is the gas density (particle type 0 in IllustrisTNG). We chose this field because of our interest in applying the method to kinetic Sunyaev-Zeldovich science in the future (see Sec. \ref{sec:conclusion}), but for the purpose of method development the precise target of the super-resolution task is not important. Although we only use one channel of data in this work, the gas particle density (baryons without stars and black holes), we can in principle include more channels in our HR data to learn the temperature, electron density, etc. Generating more than one channel, possibly with an increased pixel resolution, is left to future work, with a discussion also in Sec.~\ref{sec:conclusion}.

\subsection{Low-resolution conditional data}
\label{sec:lr_data}

Our goal is to quickly construct new HR matter fields from LR dark matter fields as a conditional for our diffusion model. With \texttt{Arepo}, we simulate dark matter fields to create the LR conditional training and LR test data. Our conditional fields are LR compared to TNG300 in three ways: they are dark matter only simulations with no baryonic physics, they contain far fewer particles, and they run across fewer time steps.

To simulate the LR training data, we use the same \texttt{NGenIC} initial seed, box size, and cosmological parameters as TNG300 described above, but now with $128^3$ dark matter particles, computing about $\sim2000$ time steps from $z=127$ to $z=0.01$. This simulation runs in about $10$ CPU hours. The HR-LR training pairs are constructed by randomly cropping $16000$ cubes of size $48^3$~px $(17.1\ \text{Mpc}/h)^3$ out of the full $264^3$~px fields, where 18\% of the volume has been set aside as a validation set. Respecting the rotational symmetry of the super-resolution operation, we also give each pair of cubes a random $\pi/2$ rotation and random chance of being mirrored. While the amount of training data is somewhat limited, we do not find evidence of over-training or memory of phases in our tests of sample diversity in Sec. \ref{sec:results_variety}. We also checked that our model did not perform better on the LR training data than on LR test data with various statistics.

For the LR test data, we run this simulation again, but now with a different \texttt{NGenIC} initial seed. We make two sets of LR test data. For the main result of this paper, presented in Sec.~\ref{sec:results}, the LR field is a $410\ \text{Mpc}/h$ length box, having 8 times the volume of TNG300; therefore, we proportionally increase to $256^3$ dark matter particles as to have the same particle density as the LR training data. For Sec.~\ref{sec:results_variety} and Appendix~\ref{sec:variable_cosmos}, we simulate LR fields at $205\ \text{Mpc}/h$. There is no HR truth for our LR test data, and so we will use summary statistics from the TNG300 training data as a truth comparison.

\section{Conditional diffusion model approach}
\label{sec:diffusion_model}

Our approach is based on the image-to-image Palette diffusion model \cite{ho2020denoising,palette}, which is capable of performing outpainting to generate larger images. We modify the model to 3 dimensions and train it to be conditional on our 3-dimensional LR simulation.

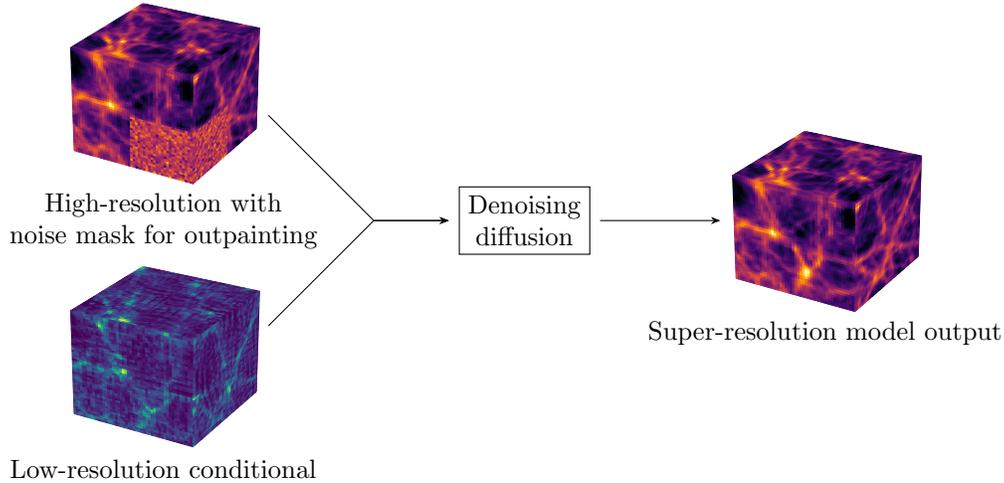
\begin{figure}
\centering
\begin{tikzpicture}
\node (lr) {\includegraphics[width=.2\textwidth]{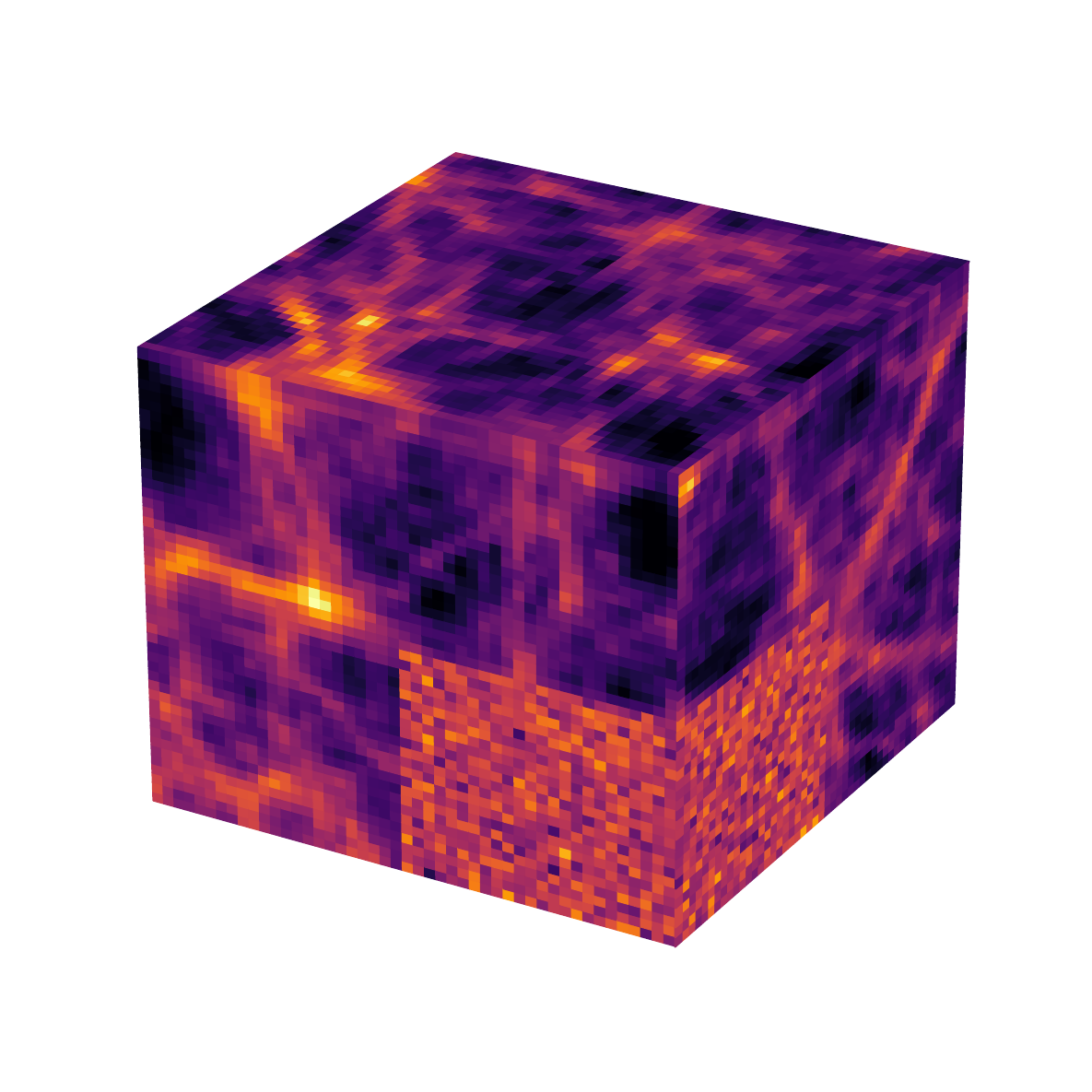}};
\node (srn) [below of=lr, yshift=-2.5cm] {\includegraphics[width=.2\textwidth]{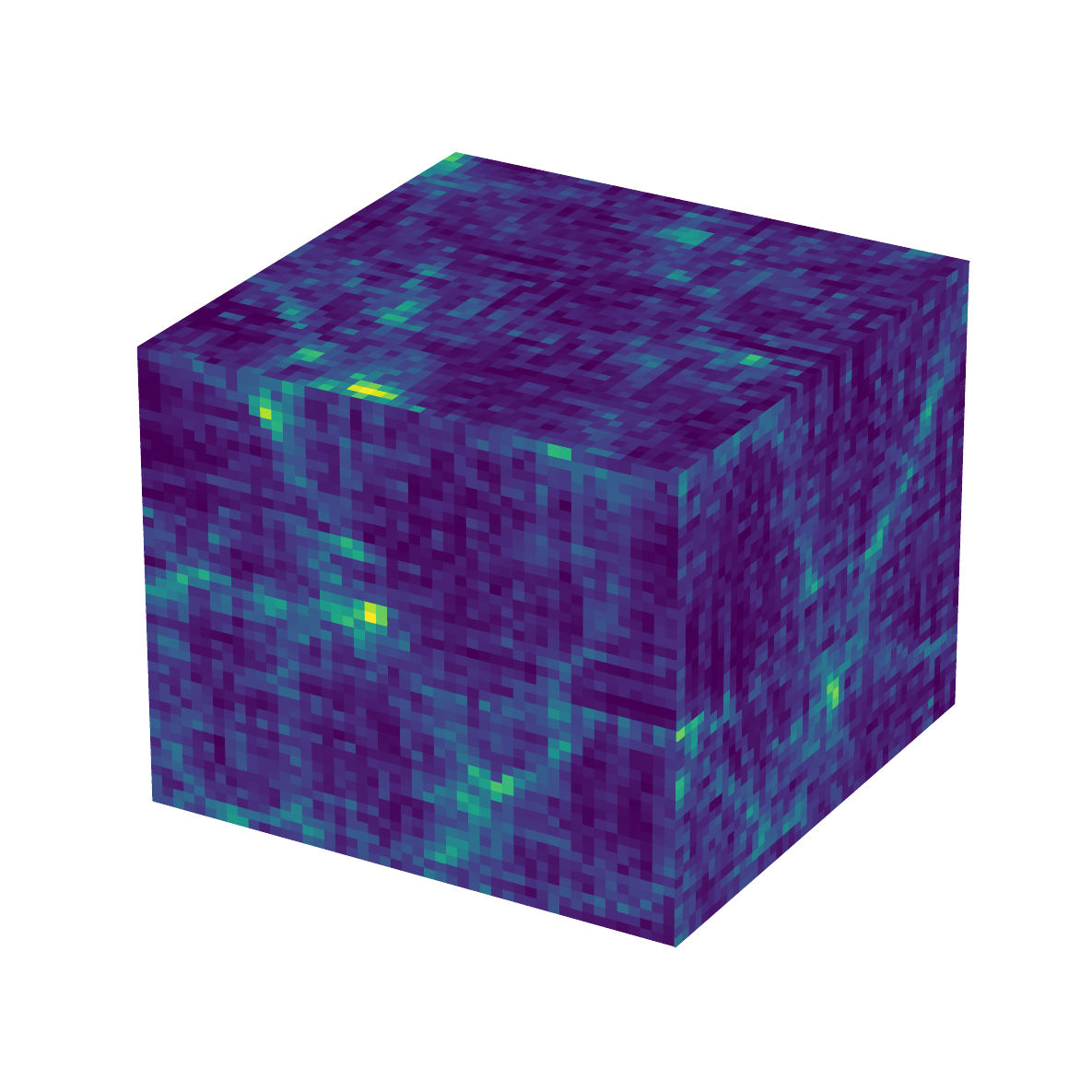}};
\node (dif) [below of=lr, yshift=-0.7cm, xshift=4.8cm, rectangle, minimum height=0.55cm, align=center, draw=black] {Denoising\\diffusion};
\node (na0) [left  of=dif, xshift= 0.12cm] {};
\node (na1) [right of=dif, xshift=-0.12cm] {};
\node (sr) [right of=dif, xshift=3cm] {\includegraphics[width=.2\textwidth]{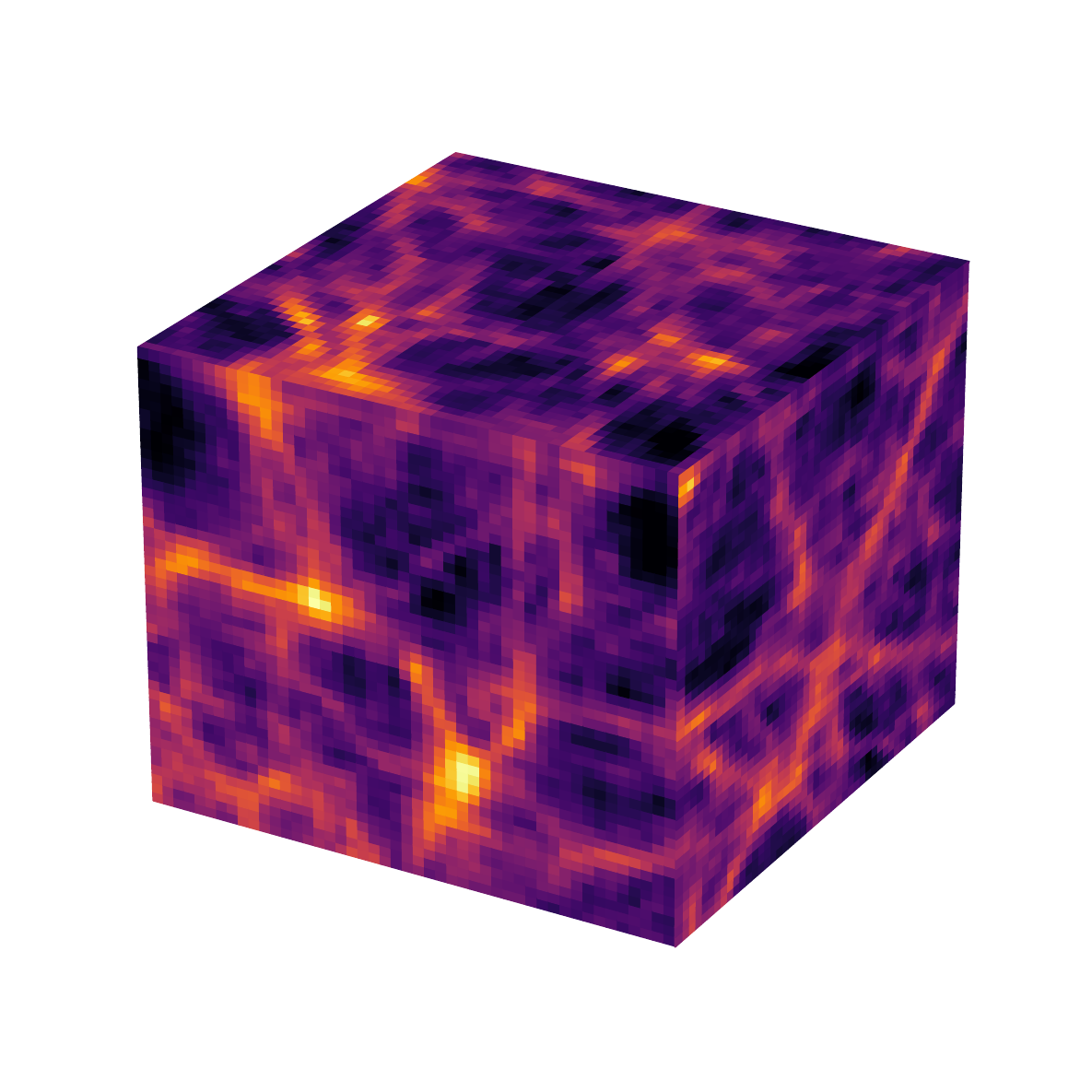}};
\node [below of=lr, yshift=-0.7cm, align=center] {High-resolution with\\noise mask for outpainting};
\node [below of=srn, yshift=-0.5cm] {Low-resolution conditional};
\node [below of=sr, yshift=-0.5cm] {Super-resolution model output};

\draw [revarrow] (na0.west) -- ++(-1cm, 0) -- ++(-1.4cm,  1.4cm);
\draw [revarrow] (na0.west) -- ++(-1cm, 0) -- ++(-1.4cm, -1.4cm);
\draw [arrow] (na1.east) -- ++(1.6cm, 0);

\end{tikzpicture}
\caption{We train a diffusion model to do outpainting on cosmological fields. By iterating this outpainting many times over, we are able to construct large SR fields given an LR conditional to guide the large scale modes.}
\label{fig:model_illustration}
\end{figure}

\subsection{Iterative outpainting to generate large fields}
\label{sec:iterative_outpainting}

\begin{figure}
    \centering
    \includegraphics[width=0.667\textwidth, trim={6.5cm, 5.5cm, 3.5cm, 5.5cm}, clip]{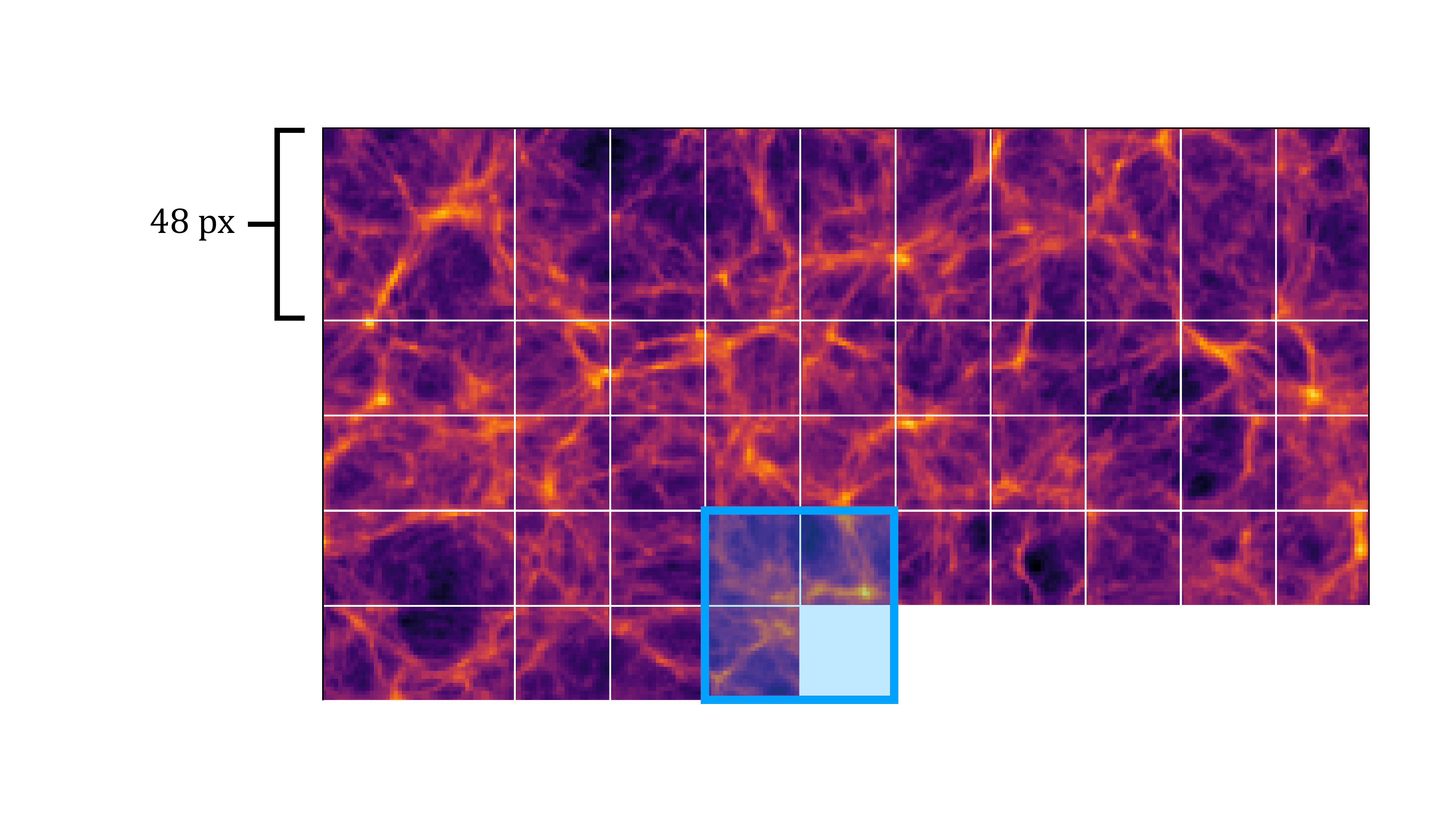}
    \caption{Iterative outpainting illustration, 2-dimensional projection. The blue cube contains the previously generated SR fields adjacent to the subsequent volume ready to be outpainted. After a new volume is generated, the blue cube then moves in row-major order for the next SR volume to be generated. After a plane of small volumes is generated to SR, the outpaintings continue in the third dimension (perpendicular to the page), and the next plane is generated, with each SR volume also conditional on the previous plane.}
    \label{fig:iterative_outpainting}
\end{figure}

This work uses massively iterative outpainting that can generate cosmological volumes much larger than the training data. Given a large LR field, an SR field is generated sequentially, patch by patch, with each new patch generated conditional on both the LR field and adjacent SR fields. The LR conditional guides the large length scales of the SR field, while the adjacent SR conditional ensures that newly generated SR fields are smooth and physically consistent with the larger SR volume. Training the model to outpaint requires masking sub-volumes of the HR data in the HR-LR pairs, as shown in~Fig.\ref{fig:model_illustration}, and thus masked sub-volumes at inference can be generated given surrounding SR data. An illustration of the iterative outpainting is shown in Fig.~\ref{fig:iterative_outpainting}. 

Our iterative outpainting method is motivated by the nature of mode coupling in the large-scale structure. On large scales, for $k\lesssim0.1\ h/\text{Mpc}$ today, the evolution of physical perturbations is linear, as modes evolve independently. On intermediate scales, $0.1\ h/\text{Mpc}\lesssim k\lesssim 0.5\ h/\text{Mpc}$ modes evolve nonlinearly but should be accurately captured by the LR dark matter N-body simulation. On smaller scales $k\gtrsim 0.5\ h/\text{Mpc}$ we want the diffusion model to model nonlinearity and baryonic feedback. This informs us about the minimum physical size required for the outpainting volumes, and the diffusion model we describe here has a fundamental mode of $k_{24\text{px}}=0.34\ h/\text{Mpc}$. The diffusion model can modify the results of the LR simulation on physical scales smaller than this scale, and can thus take into account mode coupling on these scales. Mode coupling is included by the LR simulation on large scales, but cannot be modified by the diffusion model due to its outpainting window size, and we thus assume that the LR simulation is correct on these scales. In position space, the $24$~px window corresponds to a physical length of $18.6\ \text{Mpc}/h$. This can be compared to the typical particle displacement of $5\  \text{Mpc}/h$, with an upper limit of $\sim20\ \text{Mpc}/h$~\cite{kodiramanah2020}. Most of this displacement is already included in the LR simulation, so our window size is sufficient.

It would also be possible to make large volumes without conditional patching by a different approach \footnote{We thank an anonymous reviewer for suggesting this approach.}. Since the diffusion model is convolution-based and therefore translationally invariant, by sharing the noise of adjacent volumes on the boundary, one could avoid boundary effects by construction. This is an elegant alternative that we will consider in future work. Our own approach is based on the auto-regressive way of generating data, used very successfully e.g. in PixelCNN~\cite{van2016conditional} and large language models~\cite{NEURIPS2020_1457c0d6}, which generate output sequentially. The learning task in our approach is not the same as in the alternative approach, and both sample quality and sample diversity may be different, especially with limited training data. Our approach also has the benefit that it is possible to outpaint from high-resolution data which was not generated by a diffusion model. This would be useful for example if one has a high-resolution observation of smaller parts of the sky, but only a low-resolution observation of large scales.

In detail, the iterative outpainting procedure works as follows. First, a $48^3$~px SR cube is generated conditional on a $48^3$~px LR cube; this first cube is shown in the top left of Fig.~\ref{fig:iterative_outpainting}. We move in row-major order, outpainting $24$~px at a time. The second SR volume generated is thus conditional on both its underlying $48^3$~px LR cube, as well as the $24$~px length right half of the first SR cube. After an entire plane is generated, the outpaintings move to the third dimension, with every subsequent plane conditional on the previously generated plane. The outpaintings continue in this way until the entire LR volume is generated to SR. We never break conditionality on previous adjacent SR regions, even after moving into the third dimension. Thus throughout the full SR volume, every locally outpainted volume is conditional on all adjacent previously generated volumes.

In some small regions at the outpainting boundaries, a slight discontinuity develops in the SR model output. To remedy this, we apply a linear interpolation in the $2$~px wide strip at the outpainting boundaries. This interpolation has negligible affect on the summary statistics, while making the results visually appear slightly more accurate.

A potential issue we have found is the generated data progressively slipping out of the distribution that the diffusion model was trained on. Suppose that at some early point $i$ in the outpainting chain, the diffusion model generates an out-of-distribution $\bm{y}_i+\bm{\delta}\bm{y}_i$ as a sample from $p(\bm{y}_i|\bm{x},\bm{y}_{<i})$. The next sample would be generated from $p(\bm{y}_{i+1}|\bm{x},\bm{y}_{<i},\bm{y}_i+\bm{\delta}\bm{y}_i)$, and because the model has not trained with any $\bm{y}_i+\bm{\delta}\bm{y}_i$, the out-of-distribution problem may compound. This problem arose for training the model using apparently too few model parameters. In this case, the first SR cube in the outpainting chain was too Gaussian, albeit still resembling the conditional field. However, after several more outpainting iterations, the model output quality collapsed, and appeared as very smooth blob. We overcame this problem by sufficiently increasing the number of model parameters (described below) as to never slip out of the model's learned distribution.

\begin{figure}
    \centering
    
    \begin{tabular}{cccc}
    \vspace{-0.4cm}\includegraphics[width=0.21\textwidth]{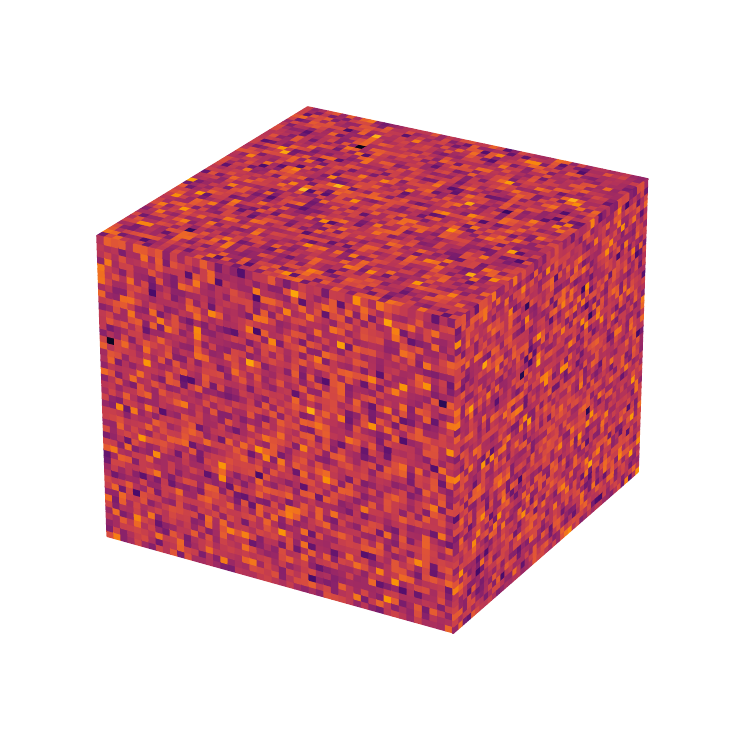} & \includegraphics[width=0.21\textwidth]{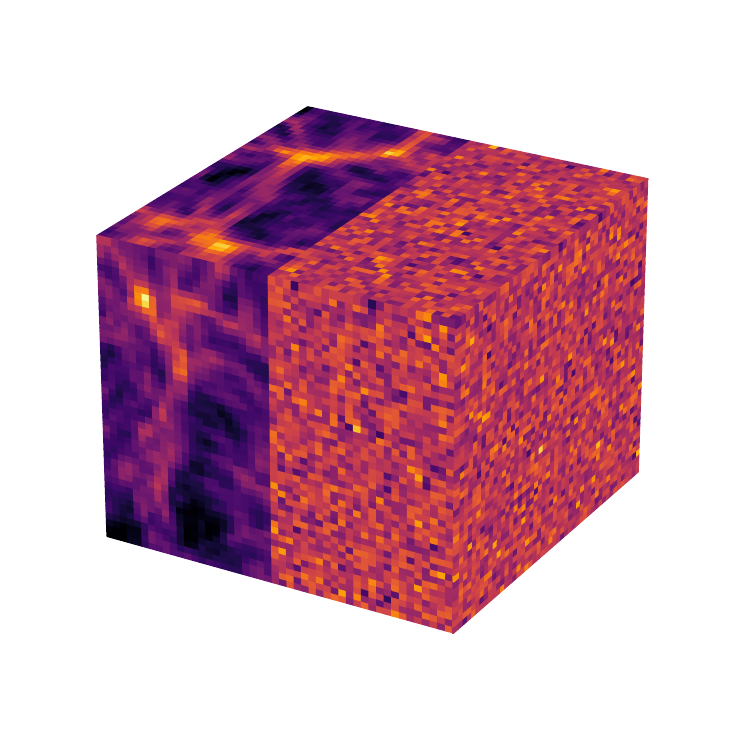} & \includegraphics[width=0.21\textwidth]{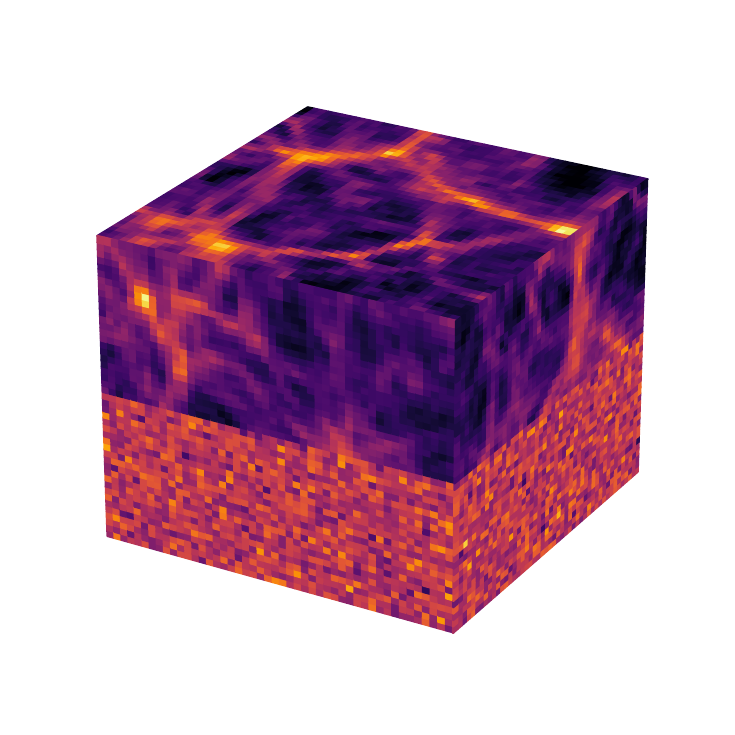} & \includegraphics[width=0.21\textwidth]{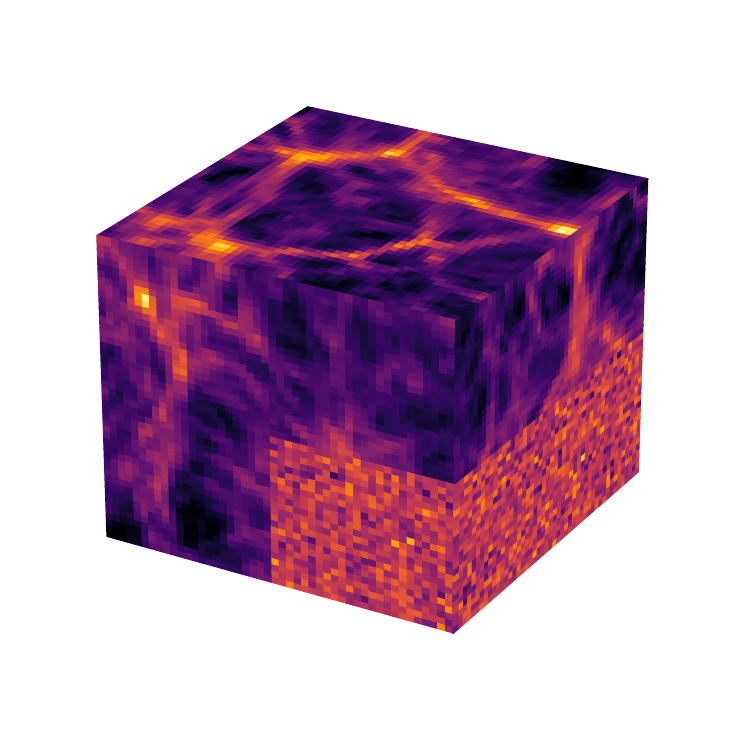}\\
    \vspace{-0.1cm}First cube & First row in & First column & First plane's\\
     & first plane & first plane & main volume\\
    \vspace{-0.4cm}\includegraphics[width=0.21\textwidth]{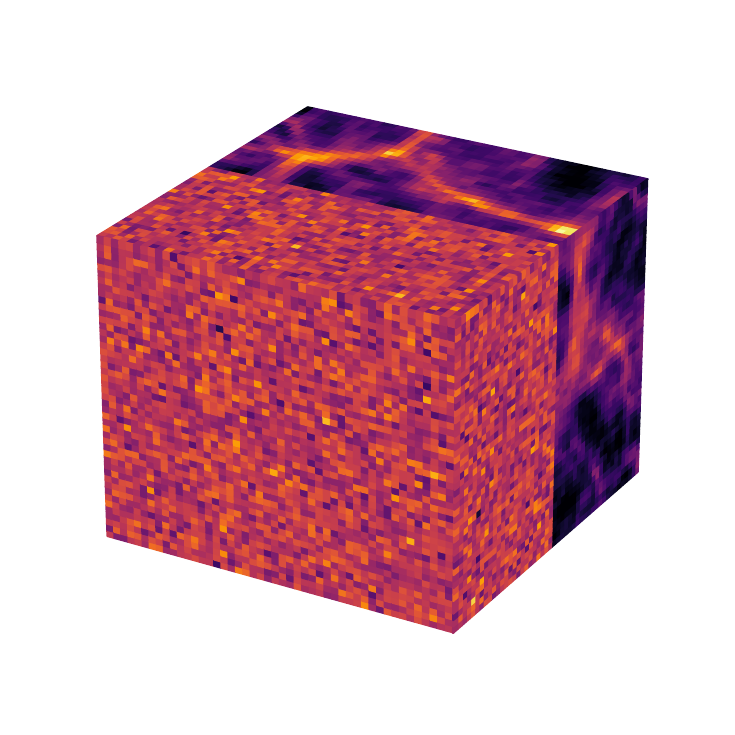} & \includegraphics[width=0.21\textwidth]{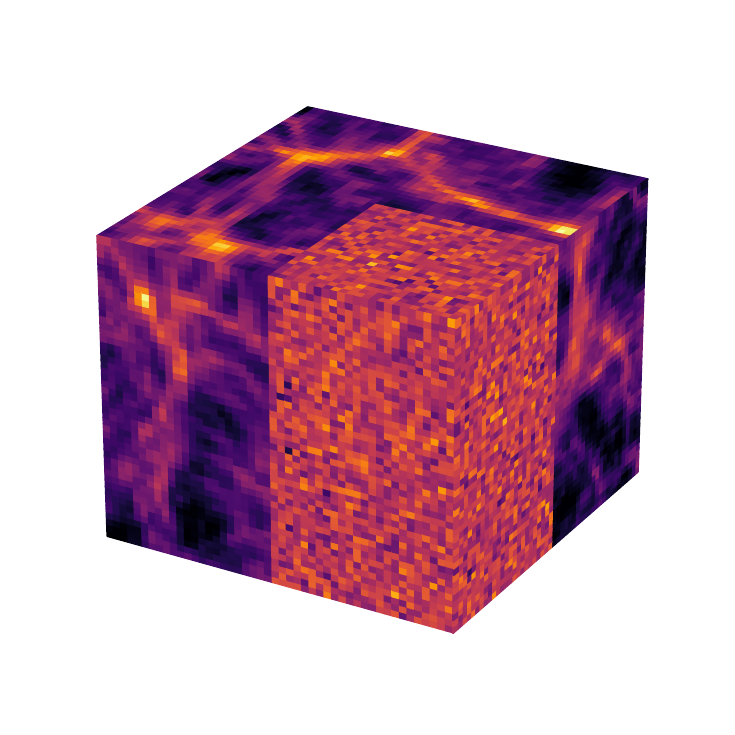} & \includegraphics[width=0.21\textwidth]{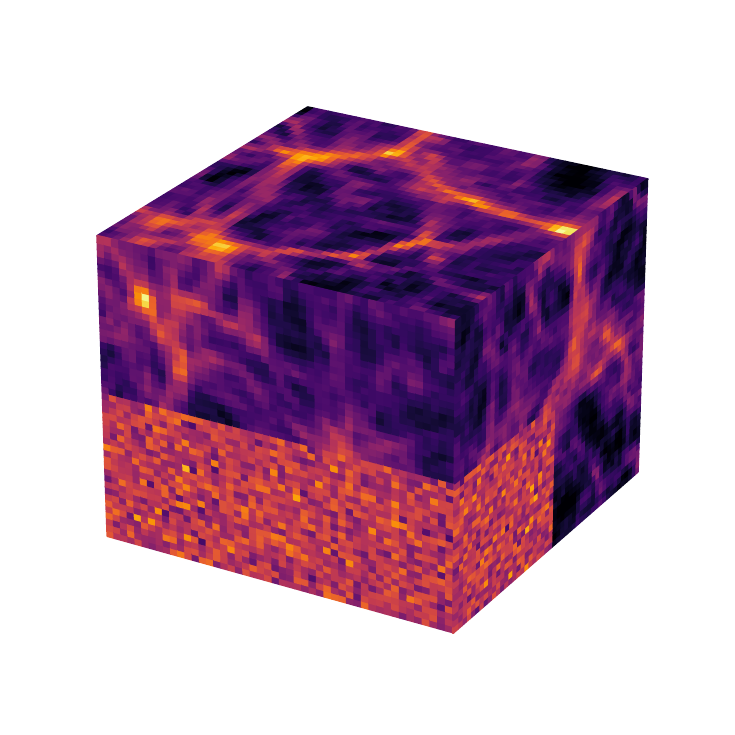} & \includegraphics[width=0.21\textwidth]{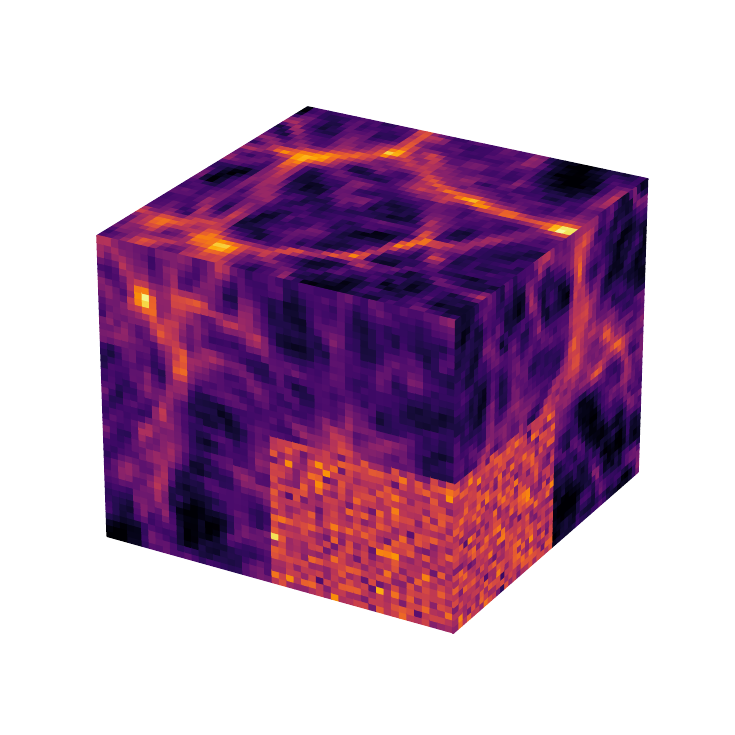}\\
    First volume in & First row in & First column in & Main volume\\
    subsequent planes & subsequent planes & subsequent planes &
    \end{tabular}
    
    \caption{There are 8 mask types (4 considering rotational symmetry) required to iteratively outpaint the full 3-dimensional volume. Each mask type is labelled as to where in the full volume it is used for outpainting. The top row of masks are used to outpaint the full first plane of cubes. The three ``subsequent planes" masks on the bottom row are for outpainting the first row and column of every plane of cubes after the first plane. The vast majority of the outpaintings belong to the ``main volume" mask (bottom right) containing noise in a single octant, used for volumes that are not located in the first plane of cubes, nor the first row or first column of any planes.}
    \label{fig:masking_demo}
\end{figure}

\subsection{Model details}
\label{sec:network_architecture}

Our model is a modified version of Palette based on code from~\cite{palette_github}. We use a U-net~\cite{Ronneberger2015} to learn the function $f_{\bm{\theta}}{(\bm{x},\bm{y}_t,\gamma_t)}$ that handles the denoising routine from $\bm{y}_t$ to $\bm{y}_{t-1}$. The details of the U-net used here are based on the ``guided diffusion" U-net described in~\cite{dhariwal2021diffusion}, which we minimally modify to operate on 3-dimensional data. A diagram of the U-net is shown in Appendix~\ref{sec:unet_figure}. The U-net takes as input $2$ channels, $\bm{y}_t$ and the conditional $\bm{x}$, and outputs $1$ channel, a prediction for $\bm{\epsilon}$. It has two downsampling steps, with the number of channels being $64$, $128$, and $256$, and similarly two upsampling steps. BigGAN residual blocks~\cite{brock2018large} are used for the downsampling and upsampling, alternating between standard residual blocks at each resolution. The middle section of our U-net has 2 residual blocks and a query-key-value attention layer~\cite{vaswani2017}. We also experimented with the SR3 U-net introduced in~\cite{sr3}, being the network used in the original Palette implementation, and we found similar model performance between these two U-nets.

Our residual blocks contain two $3\times3\times3$ convolutions, with embeddings to be dependent on the diffusion step number $t$. Prior to these convolutions is a group normalization\cite{Wu2020} with sigmoid linear unit (SiLU)~\cite{Elfwing2018} activation function. We also experimented with ReLU activation functions, with which we found similar model performance. We use $0.2$ dropout before the second convolution in each residual block. There is additionally a $1\times1\times1$ convolution in residual skip connections when the number of channels is changed in the residual block. This model has 31.5 million learnable parameters.

\subsection{Model training}

We train the diffusion model on the LR-HR pairs of $48$~px length fields described above. Due to the nature of matter clustering in cosmology, the distribution of values in both the LR and HR data is heavily skewed to the high-density tail, and so we preprocess the data by applying a natural logarithm. We want the diffusion model to take as input data values in the range $(-1, 1)$, so we additionally apply a sigmoid to the data. We should not allow the diffusion model to train on data with values well outside of $(-1,1)$, for then the $\bm{y}_T$ fields would not accurately be noisy but rather be spacially correlated with high density values in the truth $\bm{y}_0$.

After each batch of HR-LR pairs is loaded as input to the model, each HR field is randomly assigned a mask, chosen from the 8 mask types shown in Fig.~\ref{fig:masking_demo}. We train with $T=2000$ diffusion time steps with noise variances linearly increasing from $\beta_0=10^{-6}$ to $\beta_T=10^{-2}$. We use the Adam optimizer with a learning rate of $10^{-4}$, and a $0.9999$ exponential moving average learning rate decay. We train on two A100 GPUs with a batch size of $16$ per GPU, and the training loss~(Eq.~\ref{eq:loss}) converges in 70 hours. The validation loss shows no sign of over-training.

\section{Super-resolution results for a large volume}
\label{sec:results}

\begin{figure}
    \begin{center}
    \Large\textbf{Training data}\normalsize
    
    \ 
    
    \rotatebox{90}{\hspace{-1.5cm}$205\ \text{Mpc}/h$, $264$~px}\begin{tabular}{ccc}
        LR & \hspace{4.2cm} & HR\\
        \includegraphics[width=0.245\textwidth]{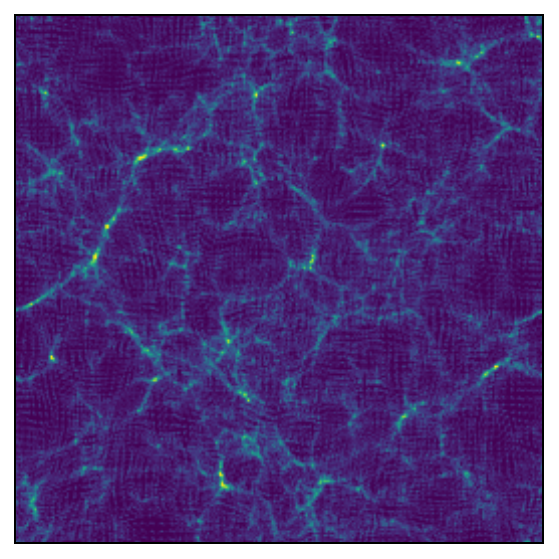} & \hspace{4.2cm} & \includegraphics[width=0.245\textwidth]{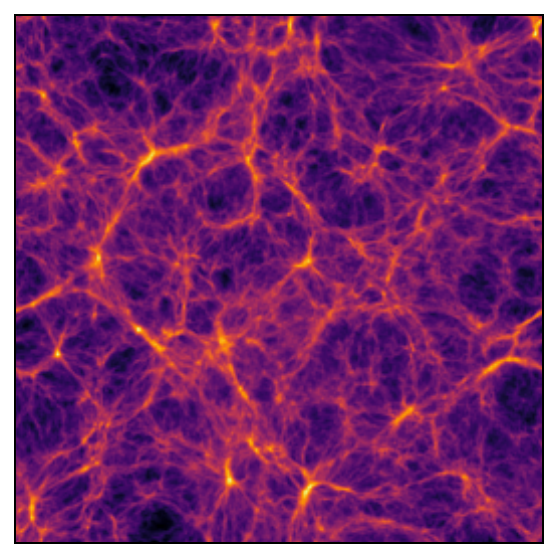}
    \end{tabular}

    \ 
    
    \ 
    
    \Large\textbf{Results}\normalsize
    
    \ 
    
    \rotatebox{90}{\hspace{-1.52cm}$410\ \text{Mpc}/h$, $528$~px}\begin{tabular}{cc}
        LR test data & SR model output\\
        \includegraphics[width=0.49\textwidth]{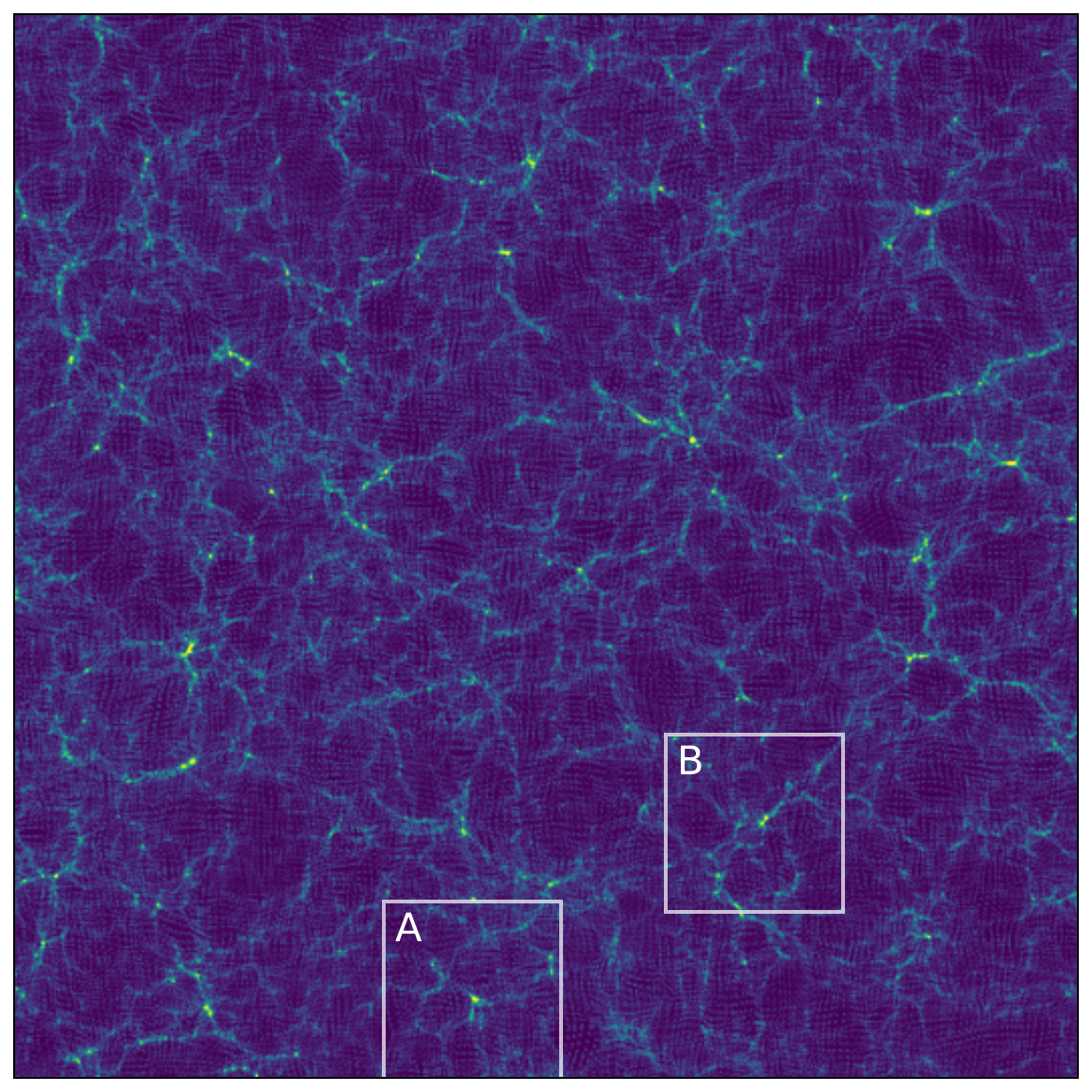} & \includegraphics[width=0.49\textwidth]{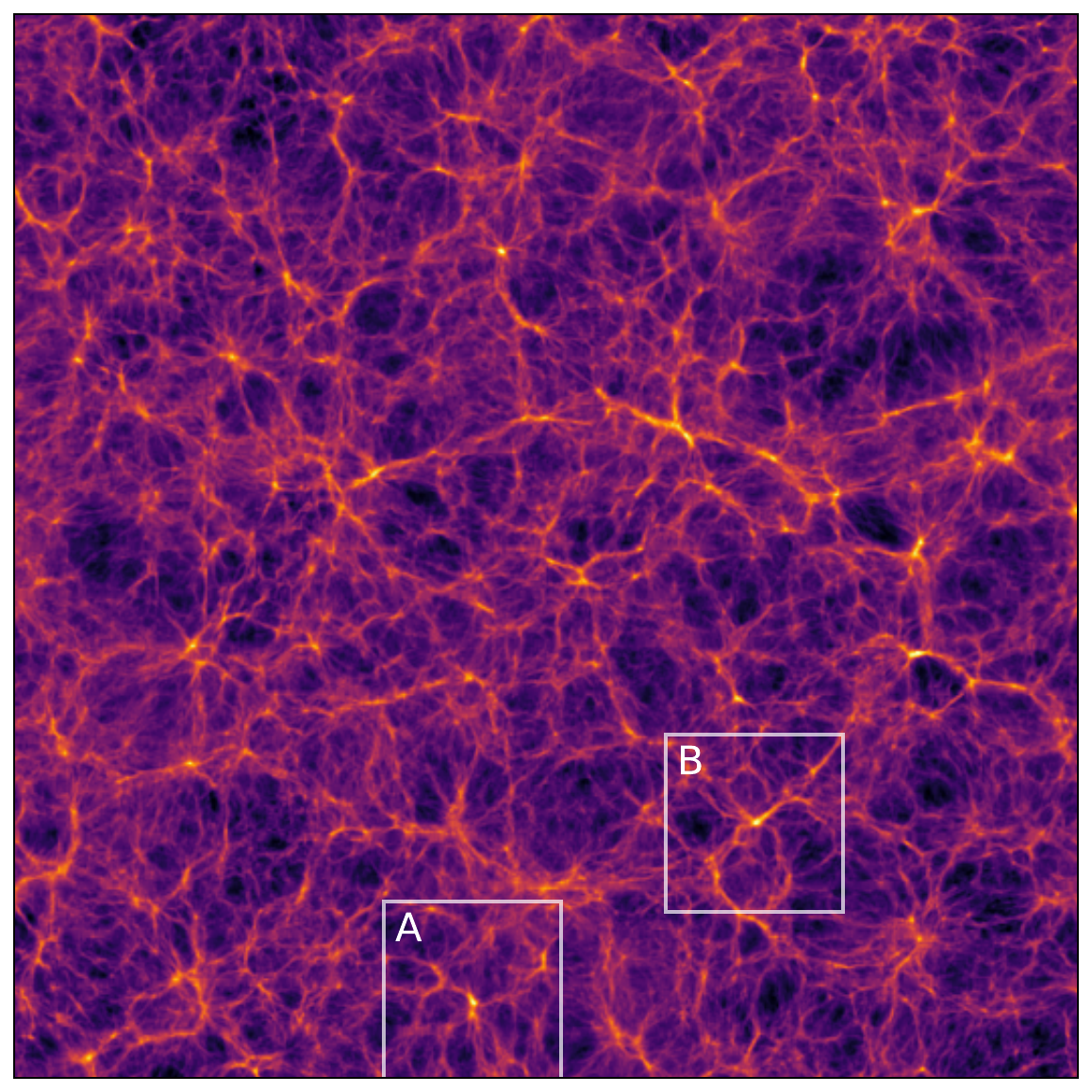}
    \end{tabular}
    %\vspace{-0cm}
    
    \rotatebox{90}{\hspace{-1.32cm}$68\ \text{Mpc}/h$, $88$~px}\begin{tabular}{cccc}
        LR zoom A & LR zoom B & SR zoom A & SR zoom B\\
        \includegraphics[width=0.245\textwidth]{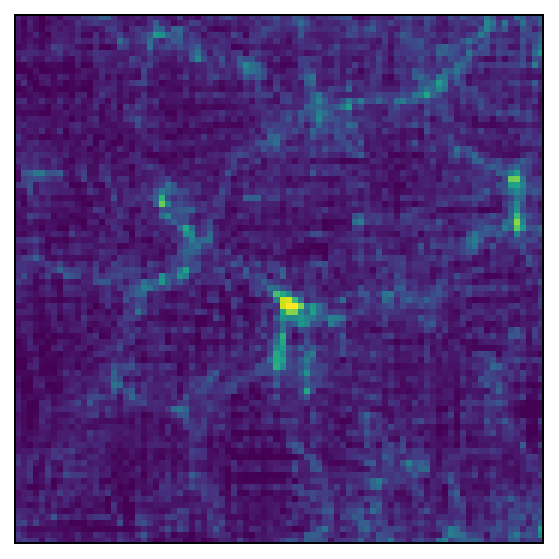} & \includegraphics[width=0.245\textwidth]{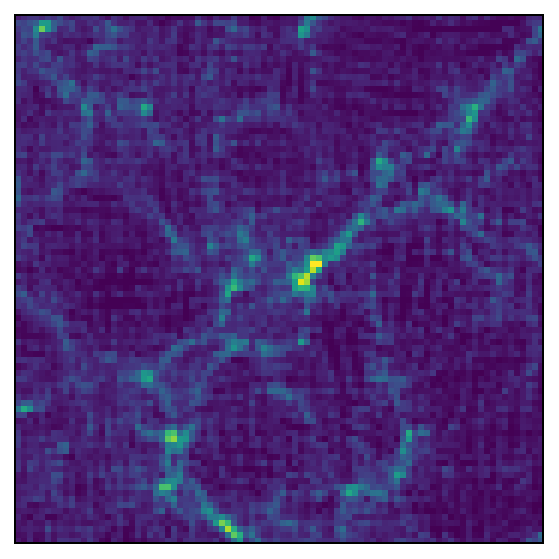} & \includegraphics[width=0.245\textwidth]{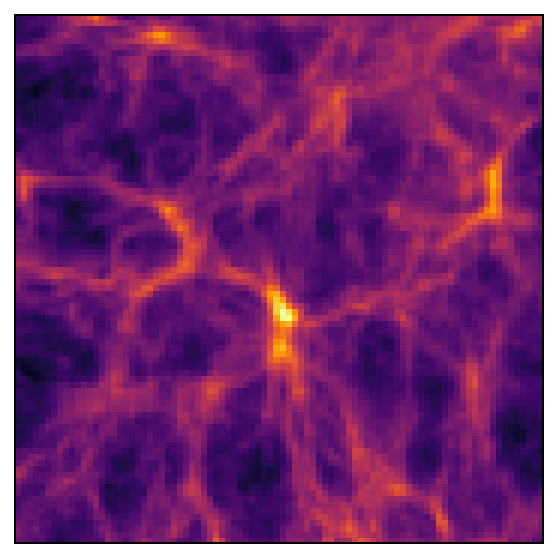} & \includegraphics[width=0.245\textwidth]{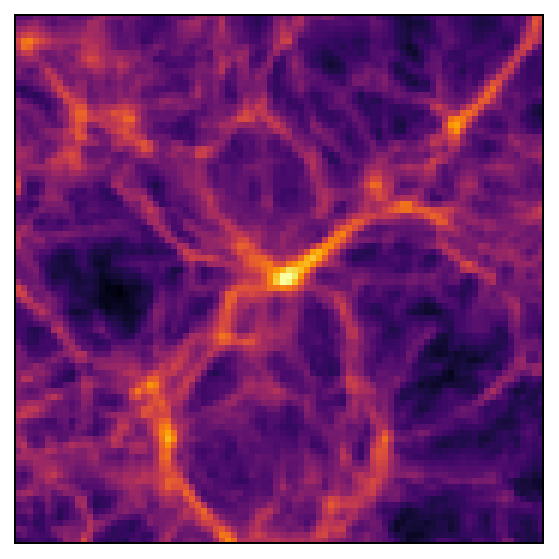}
    \end{tabular}
    \end{center}
    \caption{Matter density results for our super-resolution diffusion model in generating a volume larger than the entire training data volume. Images are 2-dimensional projections of depth $19\ \text{Mpc}/h$. (Top~row)~The training data comes from the single pair of boxes shown. The model trains on $48$~px length LR-HR pairs cut out of these boxes. (Center~left)~LR conditional test data. (Center~right)~SR model output generated with $21^3$ outpainting iterations, having 8 times the volume of the entire training data. (Bottom~row)~Two zoom-ins of the LR and SR fields.}
    \label{fig:results_visual}
\end{figure}

The main result of this work is the SR emulation of a $410\ \text{Mpc}/h\approx600\ \text{Mpc}$ length cube, with $528^3$~px, constructed with $21^3$ outpainting iterations. As the training data came from the $205\ \text{Mpc}$ length TNG300 cube meshed onto $264^3$~px, our SR result thus has an increased volume of the entire training data by a factor of 8. We show our results in Fig.~\ref{fig:results_visual}.

To generate our large SR cube, we used a linear noise schedule of $T=1250$ steps, with $\beta_0=10^{-6}$ and $\beta_T=1.5\times10^{-2}$. We found $1250$ steps to be the lowest possible before losing any measurable amount of quality in the power spectrum; a discussion of potential further speed-ups is in Sec.~\ref{sec:conclusion}. The time to generate the full cube of $21^3$ outpainting volumes was $120$ hours on a single A100 GPU.

\setlength{\tabcolsep}{0pt}

We measure the accuracy of the diffusion model with several summary statistics familiar to cosmology. Results are computed on mean zero overdensities $\delta(\bm{r})$, its Fourier transform denoted by $\delta(\bm{k})$.

\begin{figure}
    \centering
    
    \begin{tabular}{cc}
        \hspace{1cm}Probability density & \hspace{1cm}Power spectrum\\
        \includegraphics[width=0.495\textwidth]{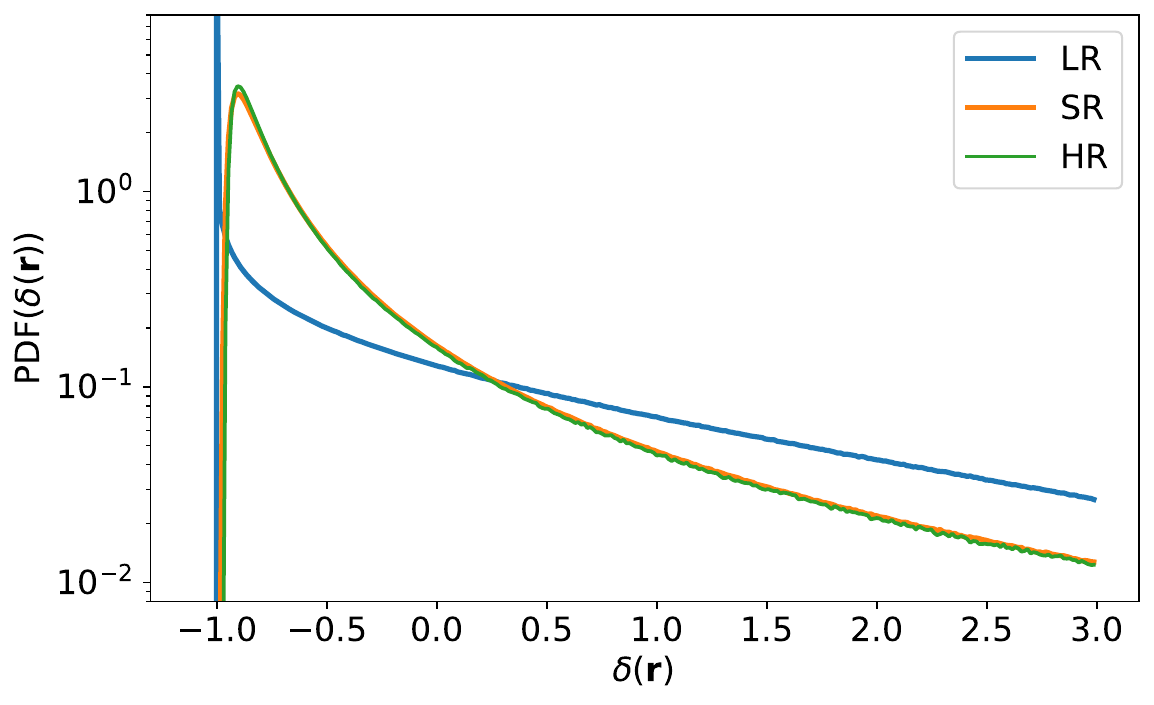} & \includegraphics[width=0.495\textwidth]{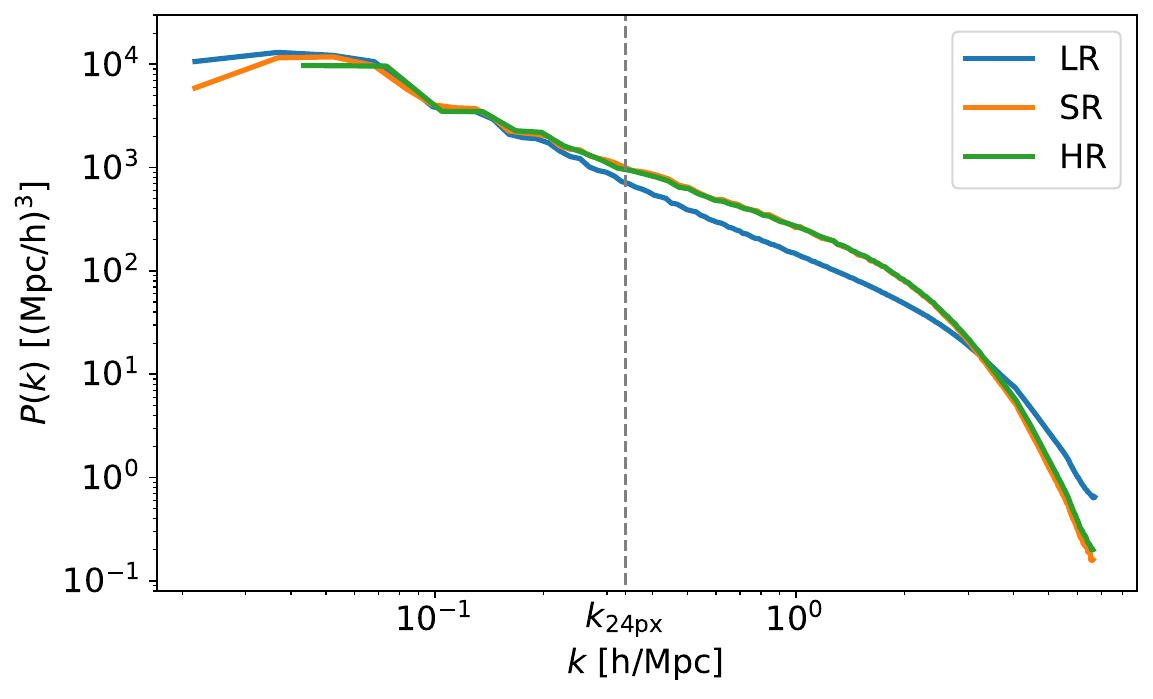}
    \end{tabular}

    \begin{tabular}{cc}
        \hspace{.85cm}Bispectrum $(k_1,k_2)=(0.15\ \frac{h}{\text{Mpc}},0.25\ \frac{h}{\text{Mpc}})$ & \hspace{.85cm}Bispectrum $(k_1,k_2)=(0.6\ \frac{h}{\text{Mpc}},1.0\ \frac{h}{\text{Mpc}})$\\
        \includegraphics[width=0.495\textwidth]{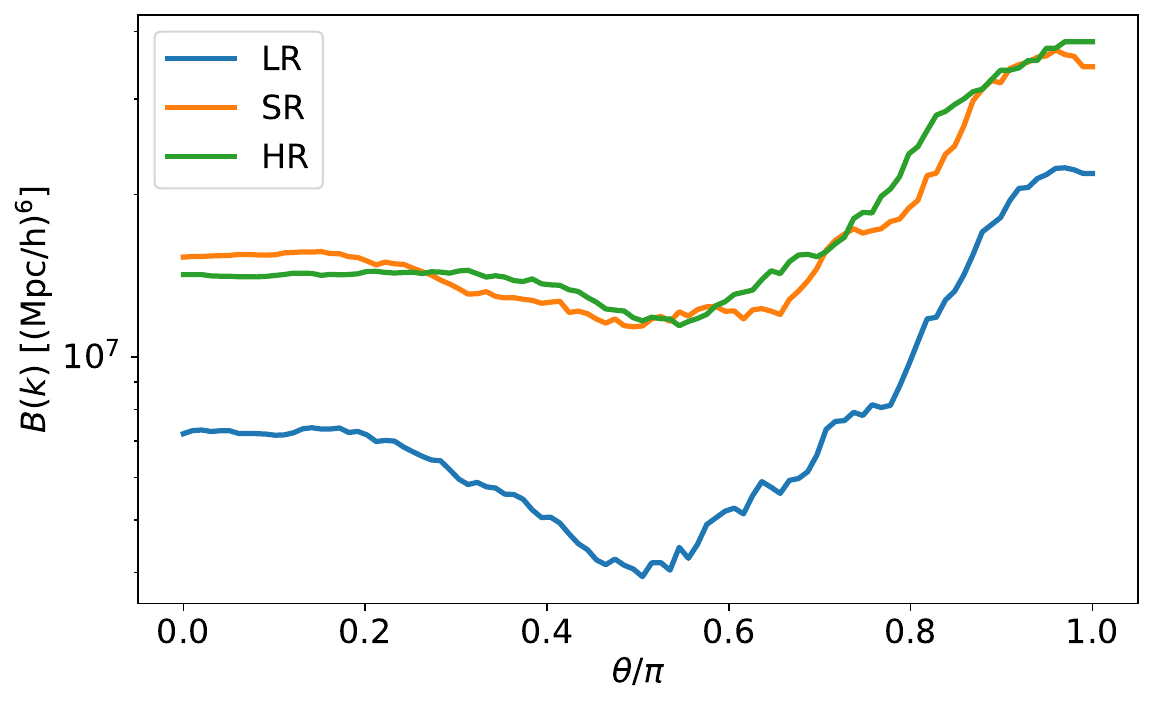} & \includegraphics[width=0.495\textwidth]{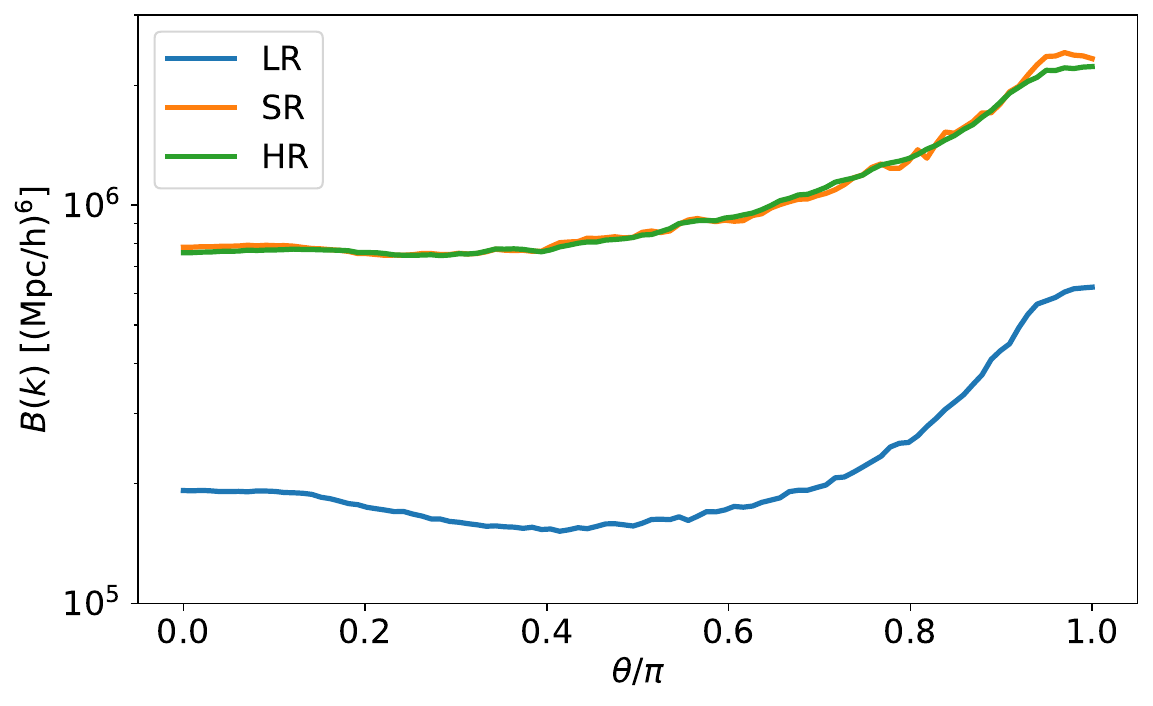}
    \end{tabular}

    \begin{tabular}{c}
        \hspace{1cm}Void size function\\
        \includegraphics[width=0.495\textwidth]{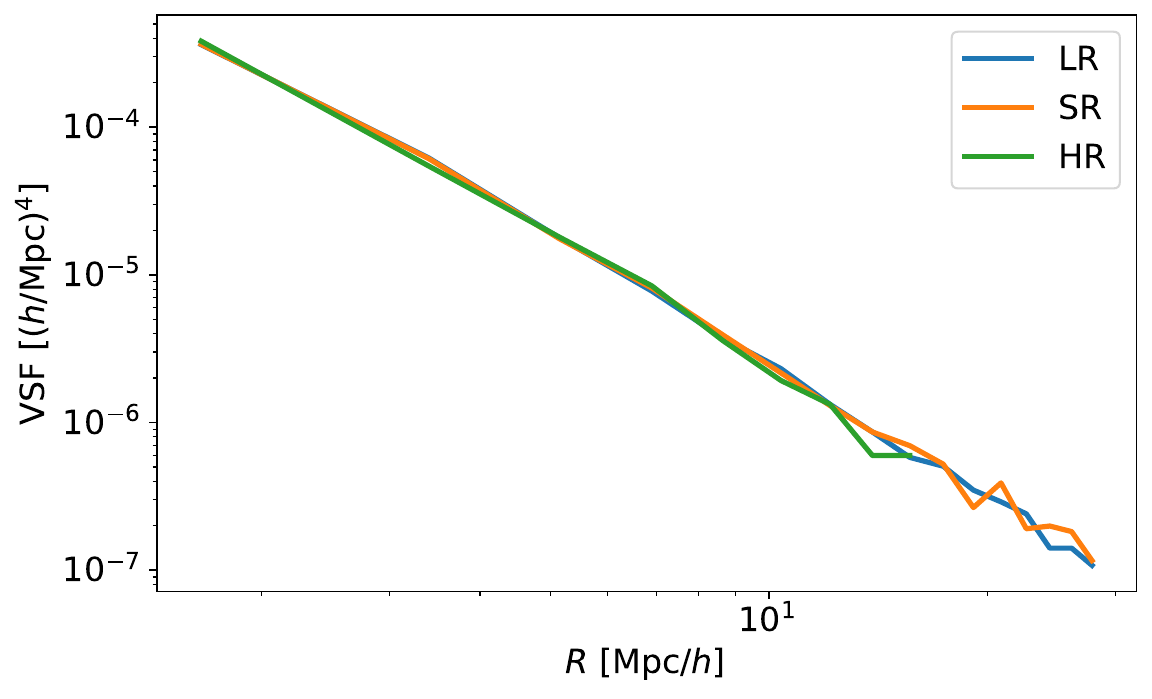}
    \end{tabular}
    
    \caption{Summary statistics comparing the $410\ \text{Mpc}/h$ length LR test field and SR model output, and the $205\ \text{Mpc}/h$ length HR training field as a truth comparison.}
    \label{fig:results_statistics}
\end{figure}

\subsubsection*{One-point probability distribution}

We plot the one-point probability density function (PDF) in Fig.~\ref{fig:results_statistics} (top left). The LR test data has many empty voxels, as it was created with $256^3$ particles in a $528^3$~px mesh. The diffusion model accurately generates SR data with the correct HR PDF, which is a smooth skew-Gaussian curve.

\subsubsection*{Power spectrum}

The power spectrum $P{(k)}$ is a two-point correlation function in Fourier space, defined by
\begin{equation}\label{eq:ps}
    (2\pi)^3P{(k)}\delta_\text{D}{(\bm{k}+\bm{k}')}=\langle\delta{(\bm{k})}\delta{(\bm{k}')}\rangle.
\end{equation}
Here $\delta_\text{D}$ is the Dirac delta function, and $\langle...\rangle$ is an ensemble average.

The power spectra comparing the LR, SR, and HR fields are plotted in Fig.~\ref{fig:results_statistics} (top right). As the LR and SR fields have twice the box length of the HR field, they have additional small $k$ Fourier modes around the $k\sim0.2h/\text{Mpc}$ turning point. We see that below the outpainting mode of $k_{24\text{px}}=0.34\ h/\text{Mpc}$, the SR power spectrum is guided by the LR power spectrum, properly emulating the baryon acoustic oscillations (BAOs) in the LR conditional. We further investigate BAOs in Appendix~\ref{sec:BAO_response}. Additionally, the SR field correctly begins to turn over for small $k$ along with the LR conditional. At high $k$, the SR simulation correctly follows the HR power spectrum, which is more suppressed for baryons than for dark matter, as physically expected. Our LR simulation dark matter power spectrum is not physically accurate at high $k$ due to its low particle resolution, but the diffusion model nevertheless learns the correct baryonic power spectrum.

A possibly surprising fact is that in the smallest $k$ power spectrum bin the SR power spectrum is somewhat suppressed with respect to the LR. In principle, the LR conditioner should dominate the power spectrum here. However, it appears that the mean density of the sequentially generated $24$~px length SR volumes can slightly drift, introducing a modulation on the very largest scales. We expect that this problem could be resolved with a multi-scale conditioner, which will be discussed in Sec. \ref{sec:conclusion}. A simpler solution would be to Fourier filter the SR map and take the largest scales directly from the LR simulation.

\subsubsection*{Bispectrum}

The power spectrum measures only the Gaussian structure, so we must also calculate the next higher moment the understand the non-Gaussian structure in our fields. The bispectrum $B{(k_1,k_2,k_3)}$ is a three-point correlation function defined by
\begin{equation}
    (2\pi)^3B{(k_1, k_2, k_3)}\delta_\text{D}{(\bm{k}_1+\bm{k}_2+\bm{k}_3)}\\=\langle\delta{(\bm{k}_1)}\delta{(\bm{k}_2)}\delta{(\bm{k}_3)}\rangle.
\end{equation}
The bispectrum is a function of three variables, so for easier analysis, we show a lower dimensional  projection of the bispectrum. We pick the first two vectors $(k_1,k_2)$ to have a constant magnitude, and plot the bispectrum as a function of the angle $\theta$ between these two vectors.

\sloppy The projected bispectrum is plotted in Fig.~\ref{fig:results_statistics} (center row) for the two choices $(k_1,k_2)=(0.15\ h/\text{Mpc},0.25\ h/\text{Mpc})$ and $(k_1,k_2)=(0.6\ h/\text{Mpc},1.0\ h/\text{Mpc})$. These two pairs of $(k_1,k_2)$ probe Fourier modes both below and above the $k_{24\text{px}}$ outpainting scale. The SR field has the same general bispectrum shape as HR for each of our two $(k_1,k_2)$ choices, with higher accuracy for the modes above $k_{24\text{px}}$. It is likely that the slight mismatch of the bispectrum on scales around $k_{24\text{px}}$ is an imperfection of our trained outpainting, while at higher $k$ the diffusion model perfectly accounts for mode-coupling. We will comment on potential improvements of the outpainting in Sec. \ref{sec:conclusion}. It could also suffice to increase the physical size of the outpainting window, which comes however at additional computational cost.

\subsubsection*{Void size function}

Voids are an important feature of the highly non-Gaussian large-scale structure at late times. The SR field should retain the overall geometry of the LR conditioner, and so we expect the quantity and size of voids in the LR, SR, and HR fields to be similar. We calculate the void size function with \texttt{Pylians}~\cite{pylians}, which uses the spherical overdensity void finder presented in~\cite{Banerjee_2016}. The void finder smooths the field with a top-hat filter of radius $R$, and then finds underdense volumes of radius $R$ below a given threshold.

The void size function is plotted in Fig.~\ref{fig:results_statistics} (bottom) for a void threshold of $0.0$ (compare to the PDF in Fig.~\ref{fig:results_statistics}). The HR void size function loses resolution at a radius of about $R=15\ \text{Mpc}/h$. The SR and LR void size functions match up to $R=28\ \text{Mpc}/h$, indicating that the SR field respects the structure of the LR field for voids larger than can be accurately measured by the full TNG300 training volume.

\section{Model output variety from a single low-resolution field}
\label{sec:results_variety}

Due to the stochastic nature of denoising diffusion models, our SR emulator is able to sample from the manifold of SR solutions $p(\bm{y}|\bm{x})$. In this section, we measure the variety of SR samples created from a single LR conditional field. Our method here may be compared to the detailed analysis in~\cite{schanz2023stochastic}, where they generate multiple SR images conditional on a single LR image, as well as multiple SR images from multiple LR images.

\begin{figure}[h]
    \centering
    
    \rotatebox{90}{\hspace{-1.52cm}$112\ \text{Mpc}/h$, $144$~px}\begin{tabular}{ccccc}
        LR & SR 1 & SR 2 & SR 3 & SR 4\\
        \includegraphics[width=0.1955\textwidth]{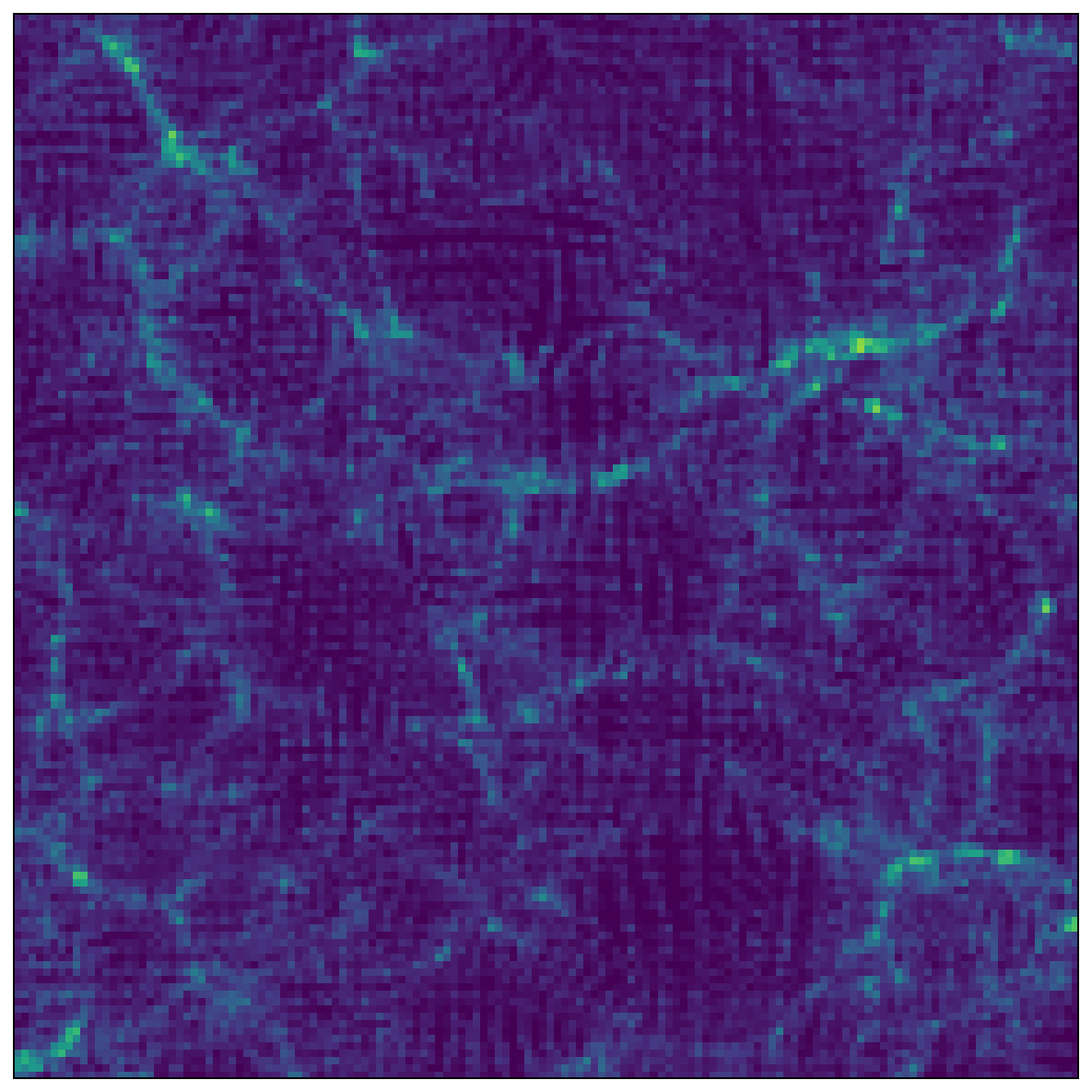} & 
        \includegraphics[width=0.1955\textwidth]{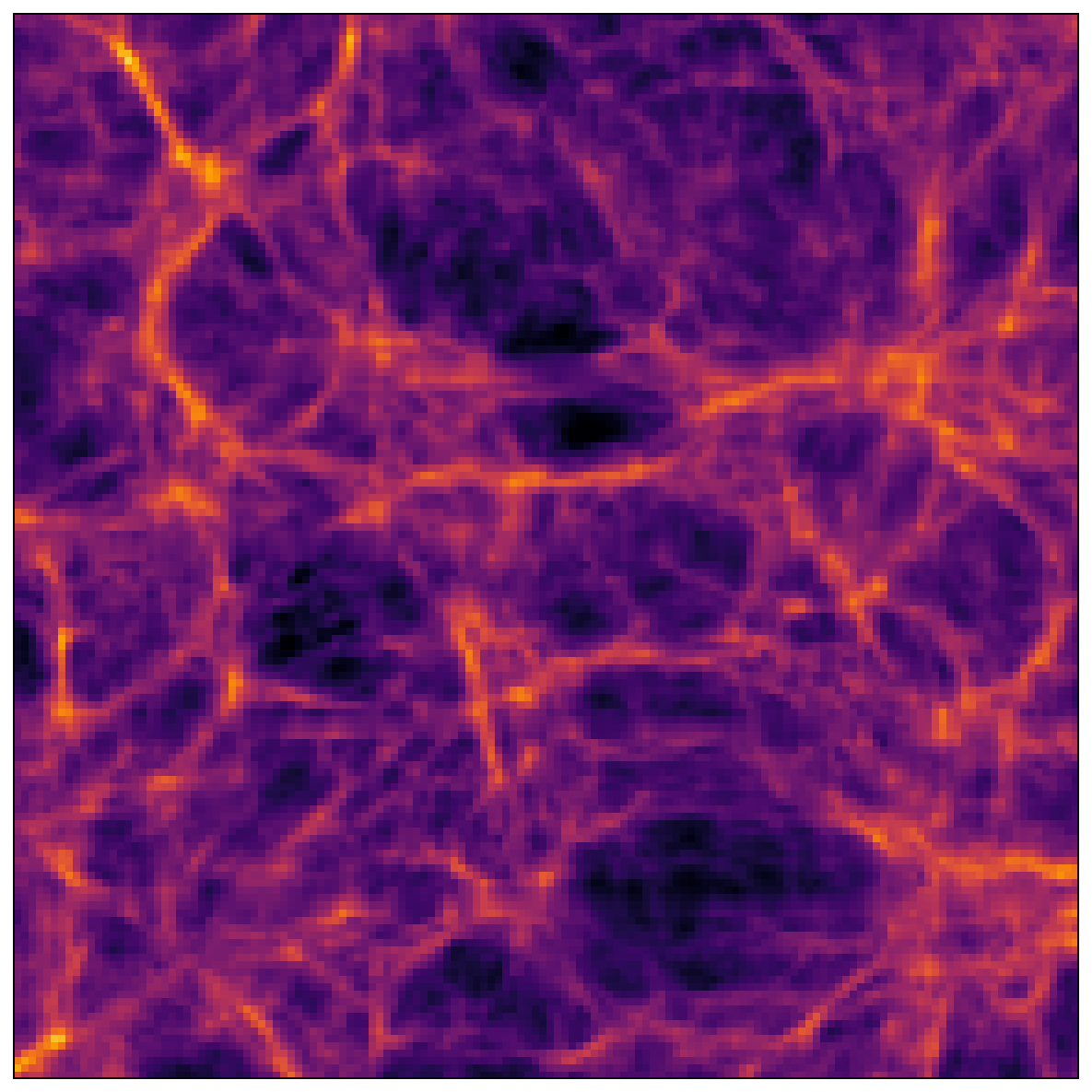} & 
        \includegraphics[width=0.1955\textwidth]{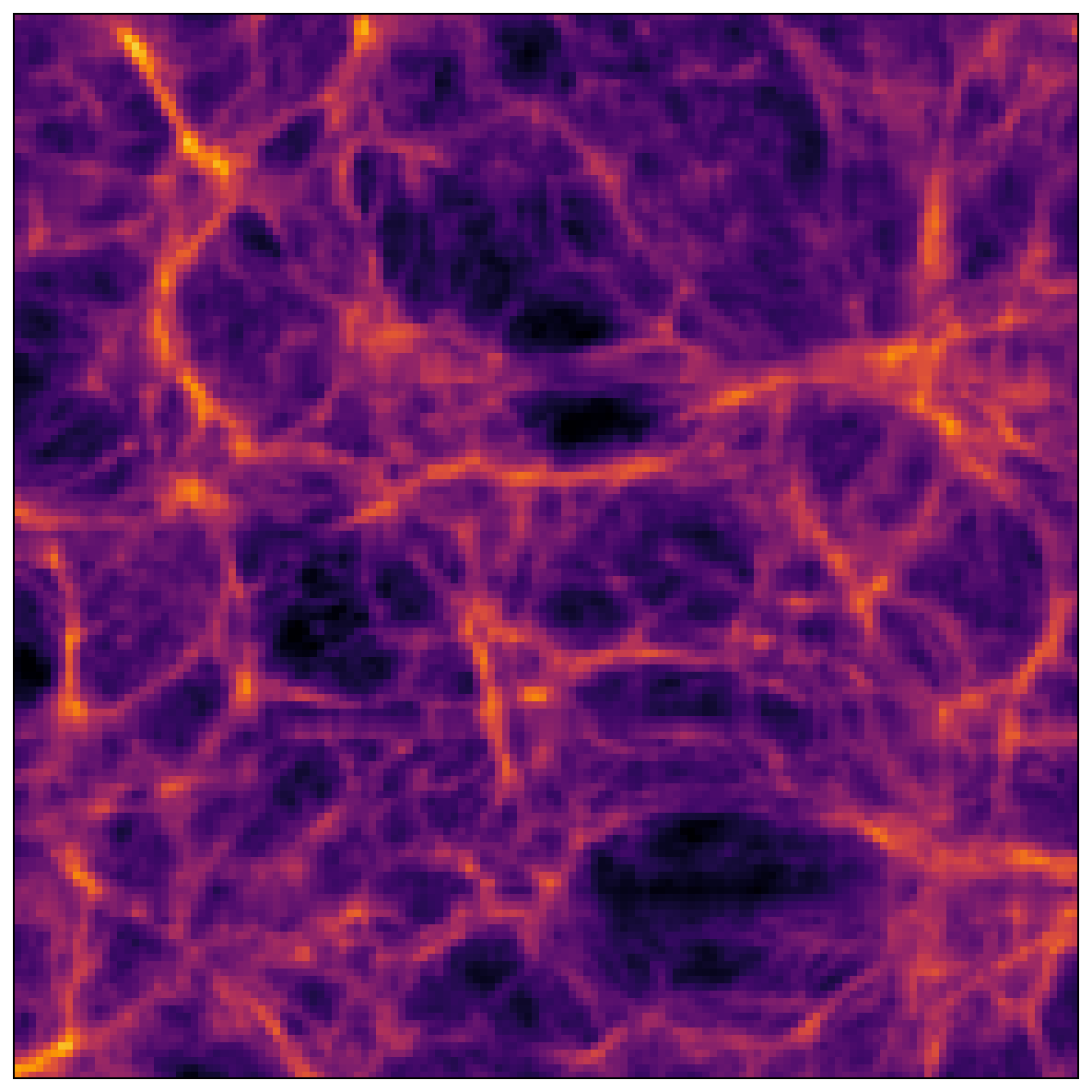} & 
        \includegraphics[width=0.1955\textwidth]{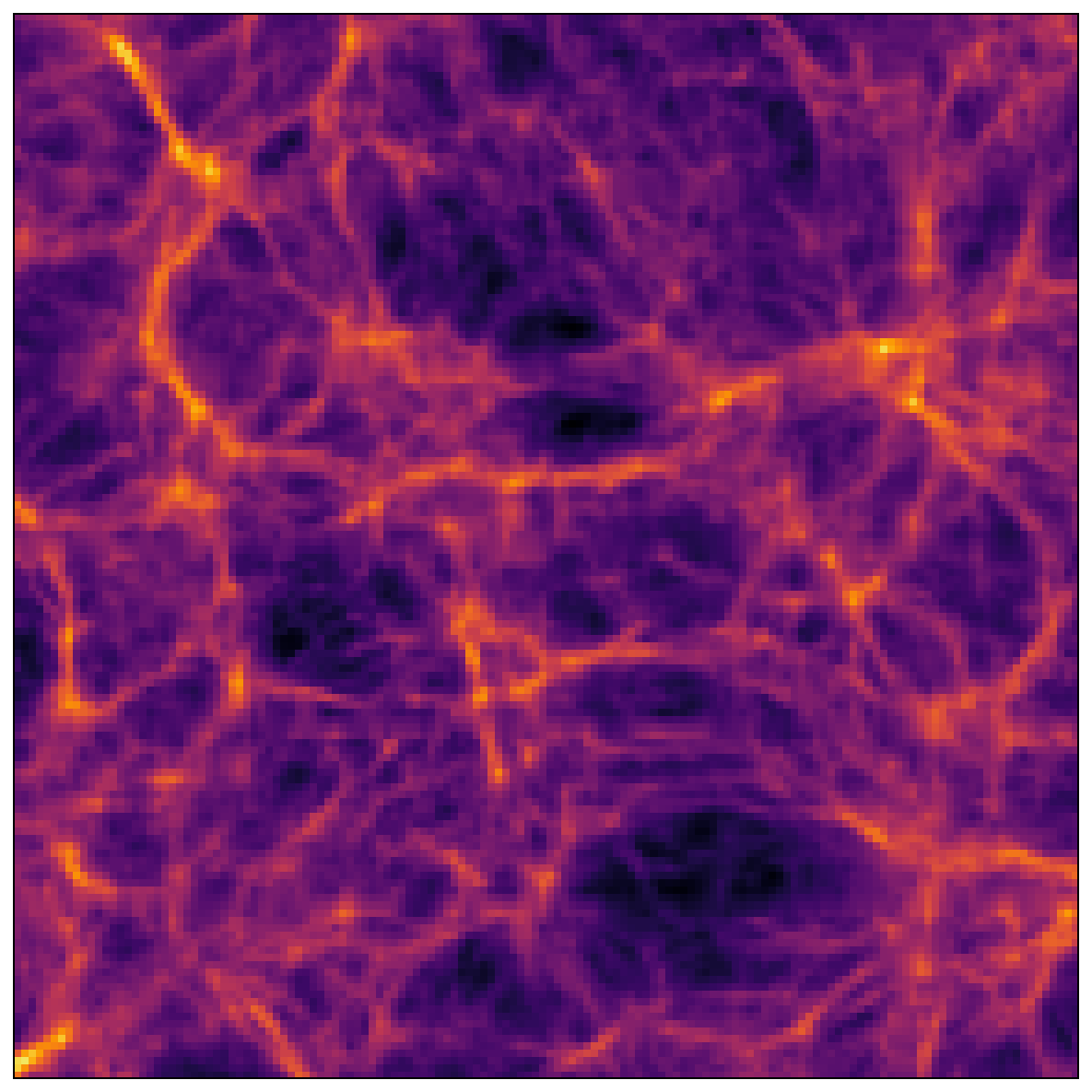} & 
        \includegraphics[width=0.1955\textwidth]{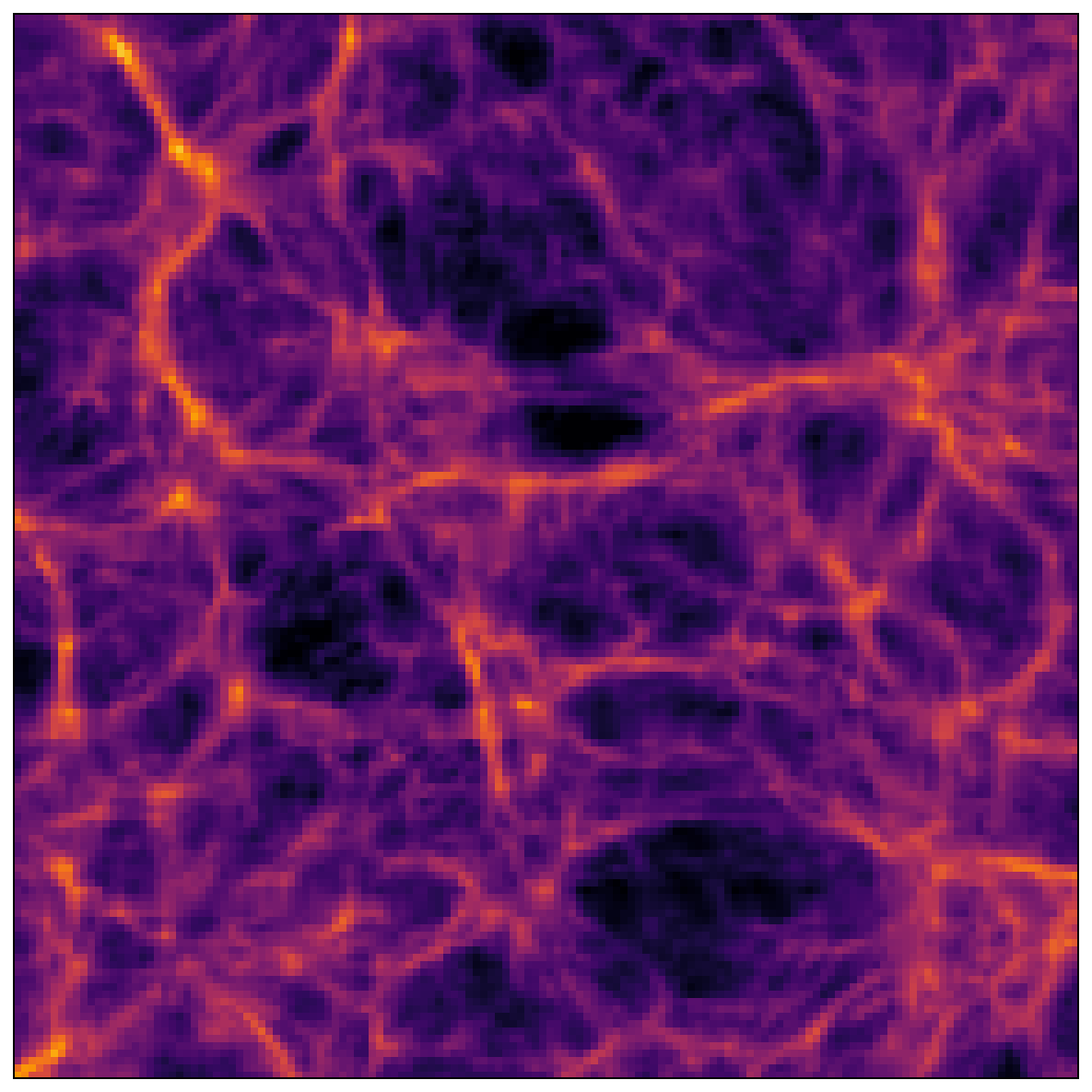}
    \end{tabular}

    \ 

    \ 

    \begin{tabular}{cc}
        SR 1 and SR 2 & SR 3 and SR 4\\
        \includegraphics[width=0.498\textwidth, trim={0 0 0 2.5cm}, clip]{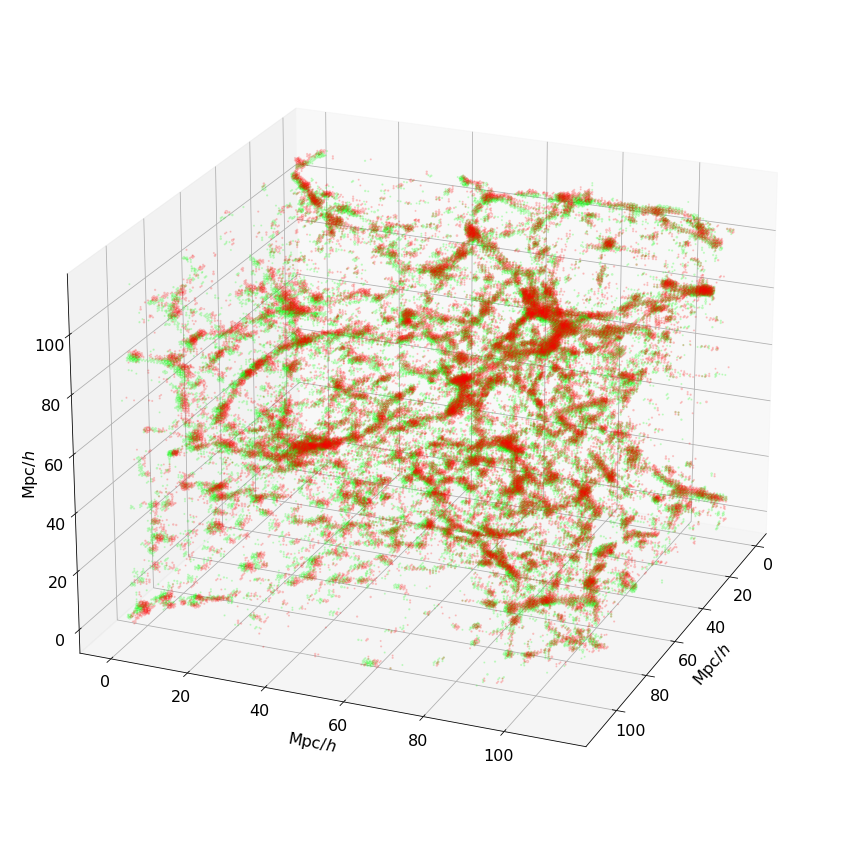} & 
        \includegraphics[width=0.498\textwidth, trim={0 0 0 2.5cm}, clip]{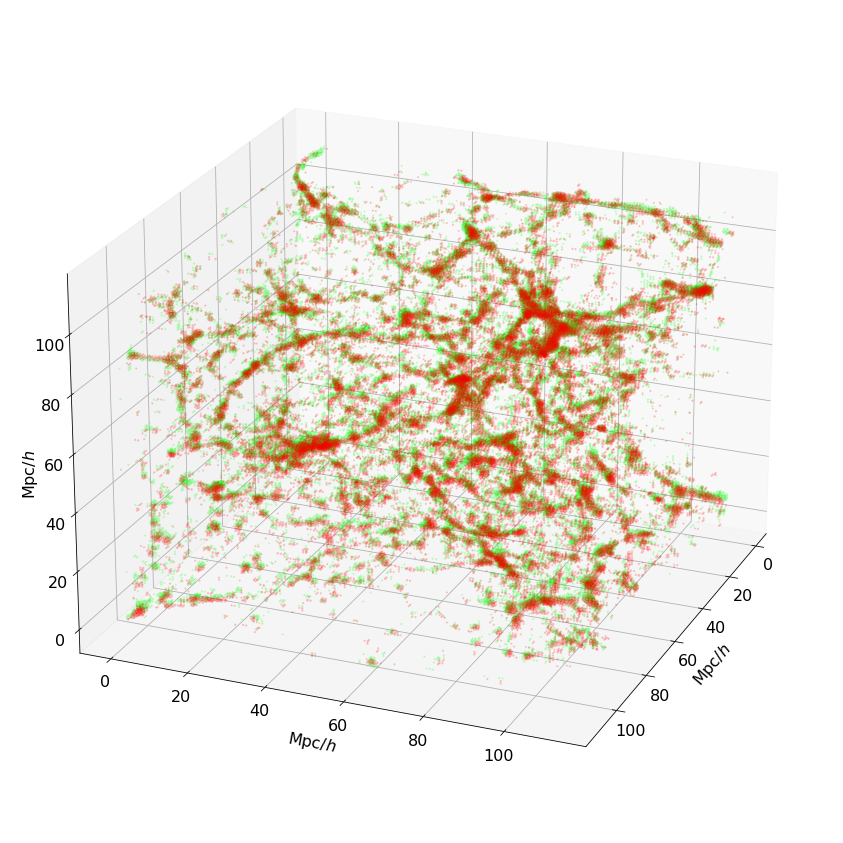}
    \end{tabular}
    
    \caption{Visual results for a variety of SR fields from a single LR conditional field. (Top row) LR and multiple SR realizations, 2-dimensional projects. (Bottom row) High density features of the above SR fields in 3 dimensions. To better visually discriminate differences between fields, we plot two different SR fields in the same volume.}
    \label{fig:results_visual_variety}
\end{figure}

We generate $25$ SR fields of size $144^3$~px (physical length $112\ \text{Mpc}/h$), each requiring $5^3$ outpaintings, conditional on the same LR field. We use the same $1250$ step noise schedule as in the previous section. Visual results are shown in Fig.~\ref{fig:results_visual_variety} for the LR volume, and four samples of the SR results. It is difficult to discern differences between the SR samples side-by-side, and so we show also two SR fields in the same 3-dimensional plot. These point plots are constructed with the large density features of two SR fields, one field in red and the other in green. When plotted on top of each other, it becomes more apparent that each SR field has a different phase in their placement and shape of halos and filaments.

\begin{figure}[h]
    \centering

    \begin{tabular}{cc}
        \hspace{1cm}Probability density & \hspace{1cm}Power spectrum\\
        \includegraphics[width=0.495\textwidth]{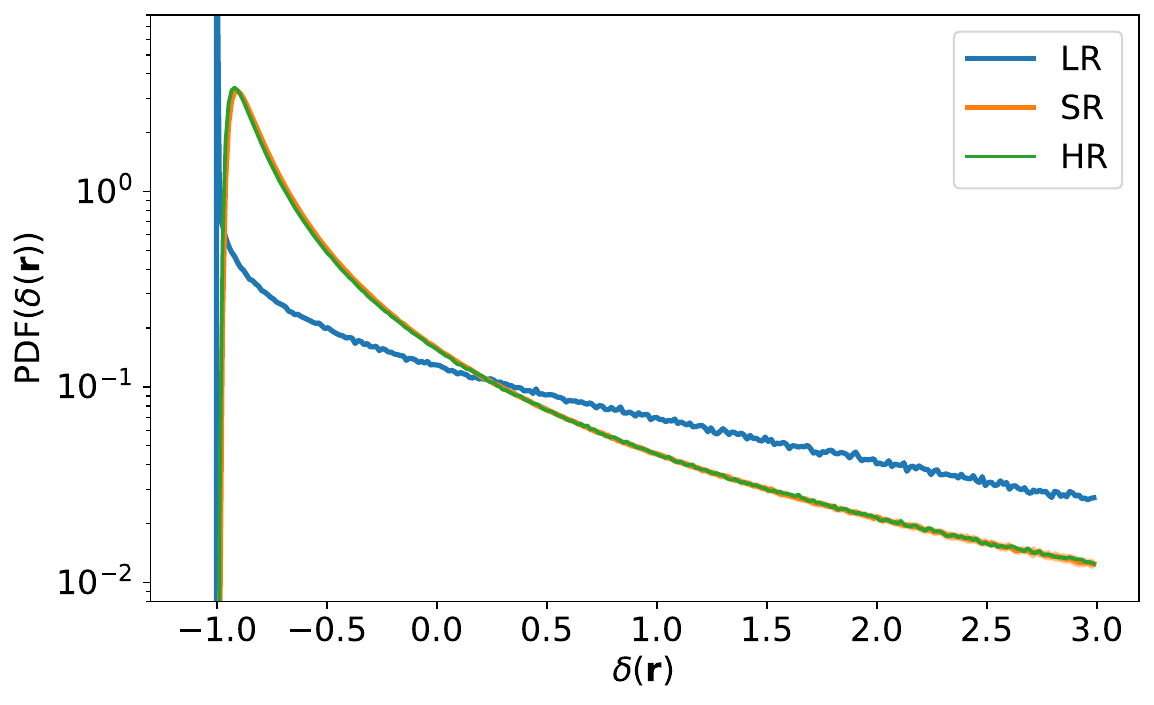} & \includegraphics[width=0.495\textwidth]{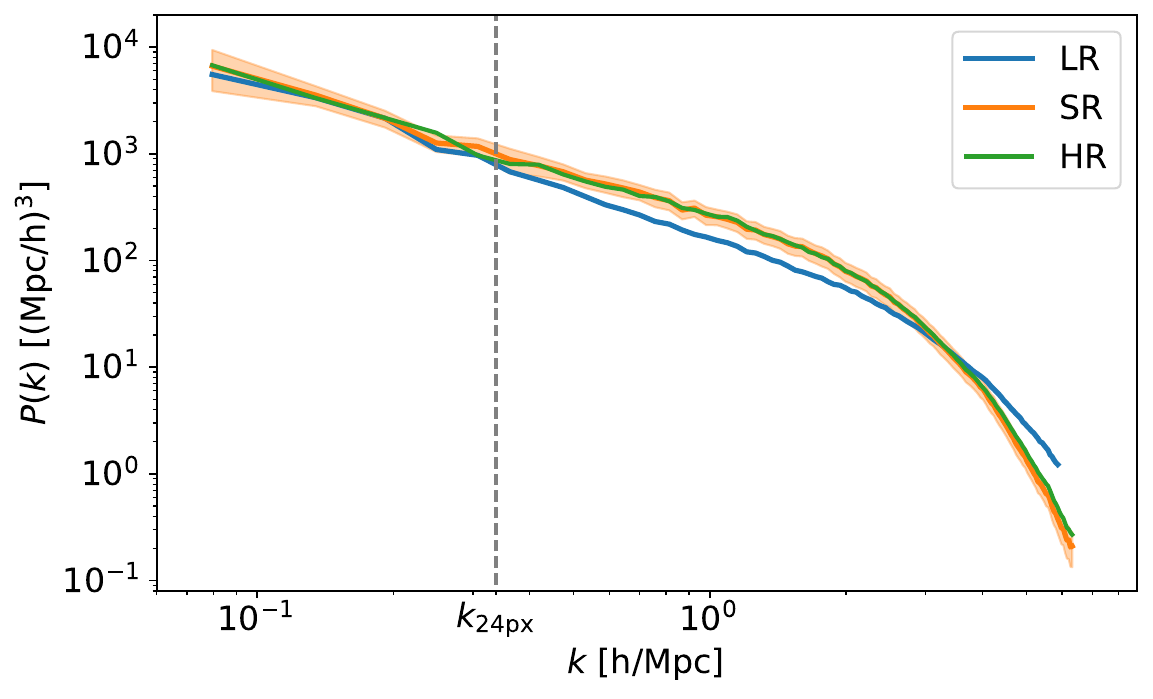}
    \end{tabular}

    \begin{tabular}{cc}
        \hspace{.88cm}Bispectrum $(k_1,k_2)=(0.6\ \frac{h}{\text{Mpc}},1.0\ \frac{h}{\text{Mpc}})$ & \hspace{.68cm}Cross-correlation coefficient between SR samples\\
        \includegraphics[width=0.495\textwidth]{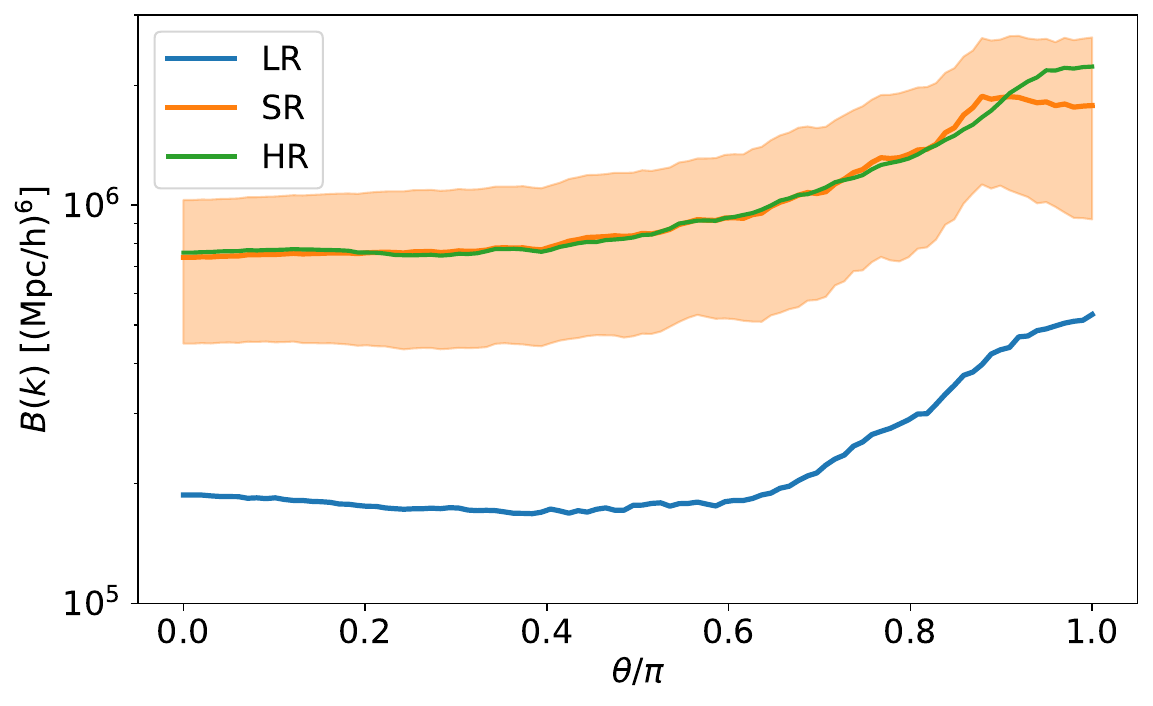} & 
        \includegraphics[width=0.495\textwidth]{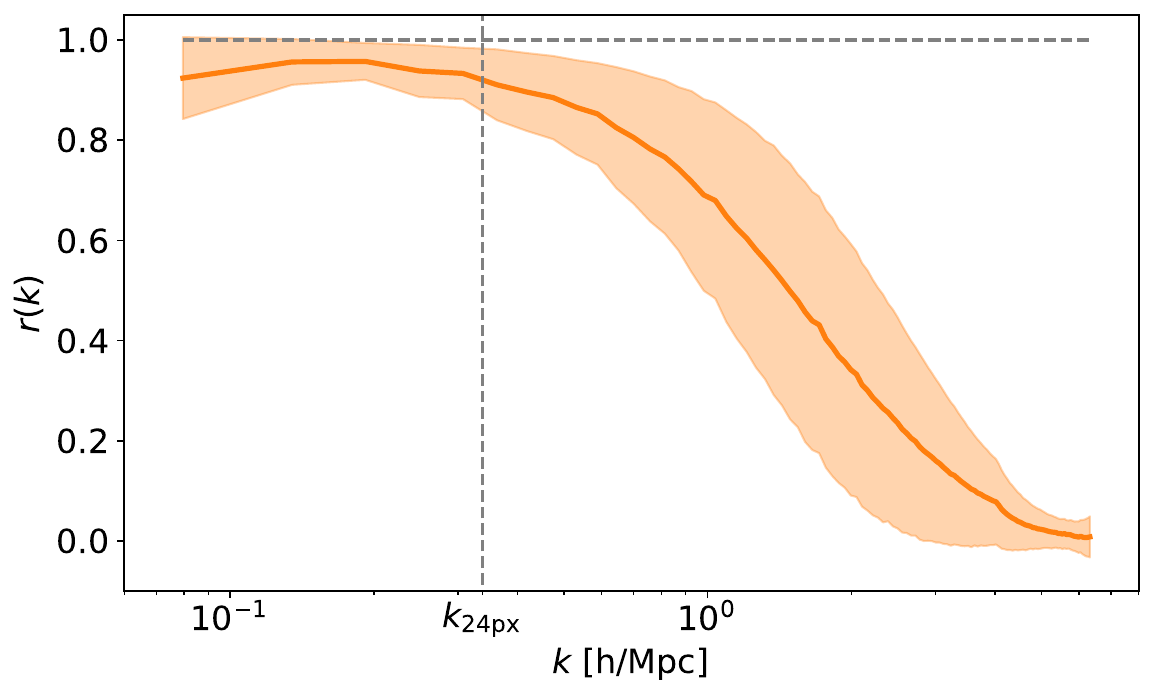}
    \end{tabular}
    
    \caption{PDF (top left), power spectrum (top right), and bispectrum (bottom left) for a variety of $25$ SR fields generated from a single LR conditional field, with $1\sigma$ bands in the SR fields. The cross-correlation coefficient of Fourier modes (bottom right) is calculated between all pairs of the 25 SR fields.}
    \label{fig:results_statistics_variety}
\end{figure}

Summary statistics for the LR and $25$ SR samples with $1\sigma$ variation between SR samples are shown in Fig.~\ref{fig:results_statistics_variety}, with the HR training data as a truth comparison. The results are highly accurate between SR and HR for the PDF (top left), power spectrum (top right), and bispectrum with $(k_1,k_2)=(0.6\ h/\text{Mpc},1.0\ h/\text{Mpc})$ (bottom left). Both the power spectrum and the bispectrum in the small SR volumes show a sample variance due to the finite mode number and the freedom of the diffusion model to modify nonlinear modes.

We further quantify the differences in the SR samples with their cross-correlations in Fourier space . For two fields $\delta_i(\bm{k})$ and $\delta_j(\bm{k})$, their cross-correlation coefficient is
\begin{equation}
    r_{ij}(k)=\frac{P_{ij}(k)}{\sqrt{P_i(k)P_j(k)}}.
\end{equation}
Here $P_{ij}(k)$ is the cross-power spectrum
\begin{equation}
    (2\pi)^3P_{ij}(k)\delta_\text{D}(\bm{k}+\bm{k}')=\langle\delta_i(\bm{k})\delta_j(\bm{k}')\rangle.
\end{equation}
We compute $r_{ij}(k)$ for all $300$ different $(i,j)$ pairs for our $25$ SR fields, and plot the result in Fig.~\ref{fig:results_statistics_variety} (bottom right). The cross-correlation coefficient is close to $1$ below the $k_{24\text{px}}$ outpainting scale as expected, as small $k$ modes are guided by the single LR field. At nonlinear scales, the cross-correlation coefficient drops to nearly $0$, signifying little correlation between SR samples. Additionally, there is almost no variance in the largest $k$-bins where the $r_{ij}(k)$ is nearly $0$, and therefore very few SR samples are correlated at the smallest length scales.

\section{Conclusion}
\label{sec:conclusion}

Super-resolution emulators are a promising tool to open the computational bottleneck of HR baryonic simulations in cosmology. In this work we evaluated the performance of a diffusion model on volumetric data, and developed a conditional outpainting scheme that can upgrade large LR volumes. We trained on data from the TNG300 simulation, and generated an SR field with 8 times the volume of the entire training data volume. Our diffusion model is capable of making accurate 3-dimensional SR emulation results, with the resulting SR volume matching the summary statistics of the training HR simulation closely. Our SR field is guided by the LR field at length scales larger than the outpainting scale. We also demonstrated the stochasticity of the diffusion model by generating a variety of SR fields conditional on a single LR field.

Leading up to this work, we found that training probabilistic generative models in 3 dimensions is not always successful, as we intended to train normalizing flows such as Real NVP~\cite{realnvp} and Glow~\cite{glow} on the same task. Normalizing flows have been recently used to learn the PDF of 2-dimensional projections of cosmological fields~\cite{rouhiainen2021,rouhiainen2022}, and our 2-dimensional super-resolution results were promising. However, the flows were not accurate in 3 spacial dimensions and we thus moved to the more expressive diffusion models. On the other hand, diffusion models are far slower at inference than normalizing flows, and this makes it challenging to generate very large volumes or many volumes. Our $410\ \text{Mpc}/h$ SR simulation took about 120 hours to generate with a single A100 GPU, and generation time scales linearly with the volume.

Further research with 3-dimensional diffusion models would be greatly aided with a faster denoising algorithm. While this work uses a conditional denoising diffusion probabilistic model (DDPM), recent developments in diffusion research include the significantly faster denoising diffusion implicit model (DDIM)~\cite{song2022denoising} and denoising probabilistic models solver (DPM-Solver)~\cite{lu2022dpmsolver}. The DDIM algorithm and DPM-Solver advertise a factor of $10$ to $100$ reduction in the number of denoising steps with a trade-off of a small loss in accuracy. We tested an implementation of DDIM in preparing this work, but our results were subpar compared to DDPM. Nevertheless, it is likely that with more work the sample generation time can be reduced by a significant factor.

It would be interesting to study different LR conditioners. Our LR conditioner is only $48$~px in field of view, which could be increased, perhaps using a multi-scale conditioner that includes up to the entire simulation. It is plausible that in this way the bispectrum at intermediate scales (see Fig. \ref{sec:results}) could be modelled even more precisely. This would also improve the modelling of SR squeezed limit N-point functions (beyond the response of the LR simulation). One may also wonder about the influence of the order of sample generation in the 3-dimensional volume. Our sample generation process breaks spatial homogeneity in principle, since we need to start generating fields somewhere in space. On the other hand, small-scale structure at some point in space is independent of small-scale structure far away from it, due to the locality of structure formation. An alternative approach which preserves spatial homogeneity exactly is discussed in Sec.~\ref{sec:iterative_outpainting}.

A physical application for our simulations is making simulations for CMB~$\times$~LSS cross-correlation analyses~\cite{kvasiuk2023autodifferentiable}. An interesting case here is kSZ tomography~\cite{Deutsch:2017ybc,smith2018ksz}, which is sensitive to the cross-correlation of the galaxy density and the electron density $P_{ge}(k)$. As shown in~\cite{smith2018ksz}, with SO and DESI, $P_{ge}(k)$ can be probed up to about $k\sim8\ \text{Mpc}^{-1}$. This required resolution is near the Nyquist frequency of the pixelization used in this work. It would thus be important to scale up the resolution of our data for future studies, which TNG300 allows owing to its large $2500^3$ particle count.

Cross-correlation analyses in general require several correlated fields, and the kSZ tomography use case requires a two field output of electron density and galaxy density. Such fields are available in IllustrisTNG, and multiple fields could be generated by our diffusion model. The most apparent way to include multiple fields would be to add additional input and output channels to the diffusion model's U-net. While technically straightforward, this approach becomes computationally challenging, especially in 3 dimensions. Another possibility is to include a second machine learning model on top of the diffusion model output. For example, invertible mappings were recently developed~\cite{andrianomena2023invertible} between different cosmological fields in the CAMELS simulations~\cite{camels2021}. 

Our method could be applied to particle level super-resolution emulation with point cloud probabilistic diffusion models~\cite{luo2021diffusion}. This would be closer to the super-resolution emulator presented in the series of papers~\cite{li2021_1, li2021_2, li2023} or~\cite{schanz2023stochastic}. Point cloud diffusion models have been recently used to generate QCD jets with high precision~\cite{Mikuni:2023dvk,Leigh:2023toe}. For some analyses, a hybrid generator that can both make a continuous matter field and a point-cloud galaxy field would be useful.

Finally, it would be useful to be able to vary cosmological and astrophysical parameters. In Appendix~\ref{sec:variable_cosmos} we explored transferring new cosmologies with different cosmological parameters from the LR conditional field to the SR field. Unfortunately, our diffusion model did not behave well with out-of-distribution LR fields. A more straightforward implementation of varied cosmological parameters requires multiple training simulations with different values of such parameters. For the IllustrisTNG simulations, this is not currently available, but the CAMELS project and its planned extensions can be useful. For our approach, we require simulations that both include large linear and small nonlinear scales. In principle, such parameter dependence can be included in the diffusion model with techniques such as style transfer and Low-Rank Adaptation~\cite{hu2022lora}, the latter of which has been successful at concept tuning and concept fusion with text-to-image diffusion models~\cite{gu2023mixofshow}. With larger but still realistic GPU resources, it would be possible to generate a few large volume simulations with different astrophysical parameters for the same cosmology.

\section{Acknowledgements}

We thank Ying Fan for helpful discussion. This material is based upon work supported by the U.S. Department of Energy, Office of Science, Office of High Energy Physics under Award Numbers DE-SC-0023719 (G.S.) and DE-SC-0017647 (G.S.,M.M.). M.M is supported by NSF grant 2307109. Model training and inference used resources from the Data Science Institute at the University of Wisconsin-Madison.

\bibliographystyle{unsrturl}%unsrt}
\bibliography{refs}{}

\appendix

\section{SR response to BAOs in the LR conditional}
\label{sec:BAO_response}

Our diffusion model's SR output follows the basic structure of the LR power spectrum at large length scales, even though the SR field is constructed with many smaller conditional outpaintings. We can demonstrate the response of our SR field to LR modes by examining BAOs. As we saw in Sec.~\ref{sec:results}, the SR output was able to accurately emulate the BAOs around $k\sim0.1\ h/\text{Mpc}$, lower than our outpainting scale of $k_{24\text{px}}=0.34\ h/\text{Mpc}$.

\begin{figure}
    \centering
    \textbf{BAOs emulated from LR to SR}

    \ 
    
    \rotatebox{90}{\hspace{-1.52cm}$410\ \text{Mpc}/h$, $528$~px}\begin{tabular}{ccc}
        LR test data & \hspace{0.5cm} & SR model output \\
        \includegraphics[width=0.333\textwidth]{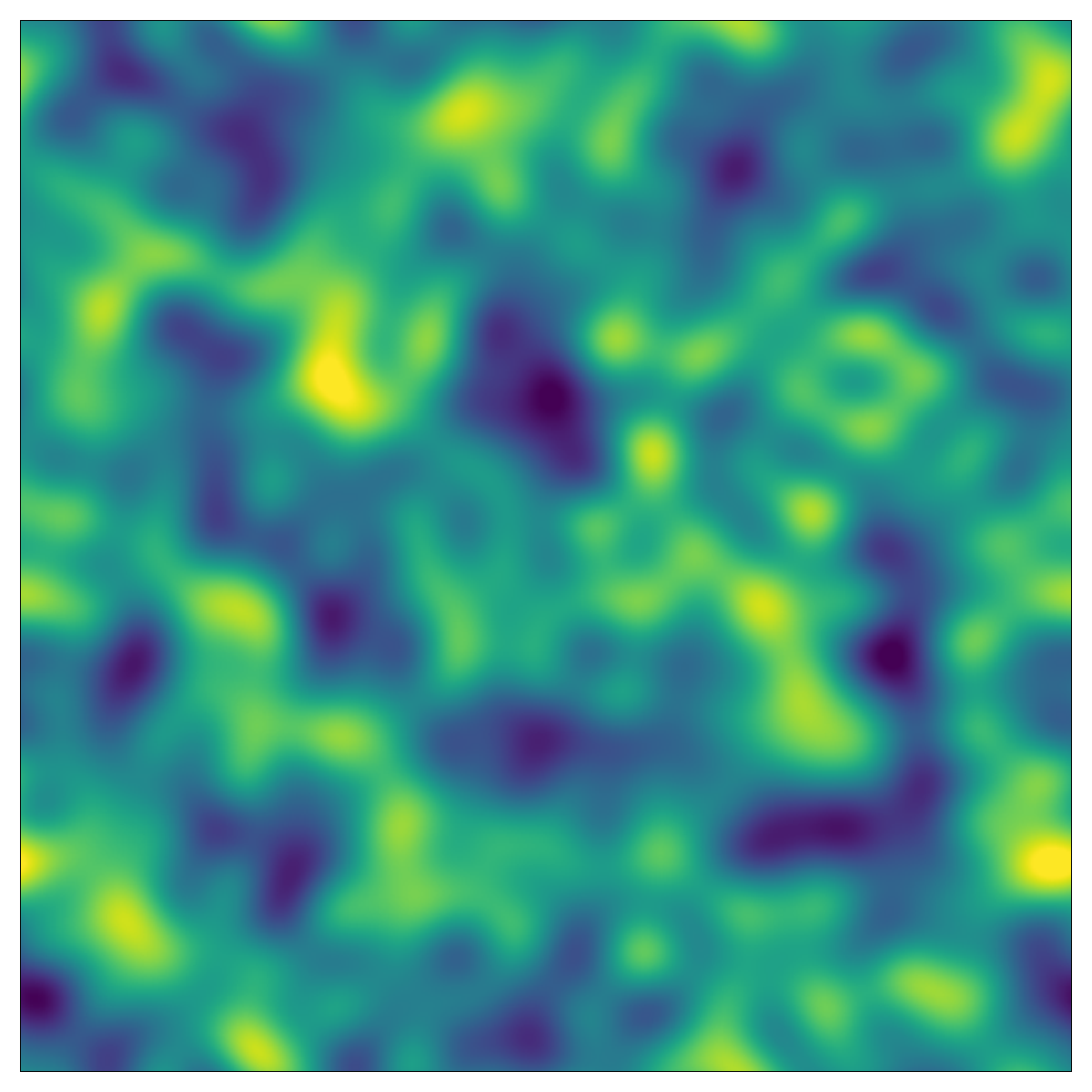} & \hspace{0.5cm} & \includegraphics[width=0.333\textwidth]{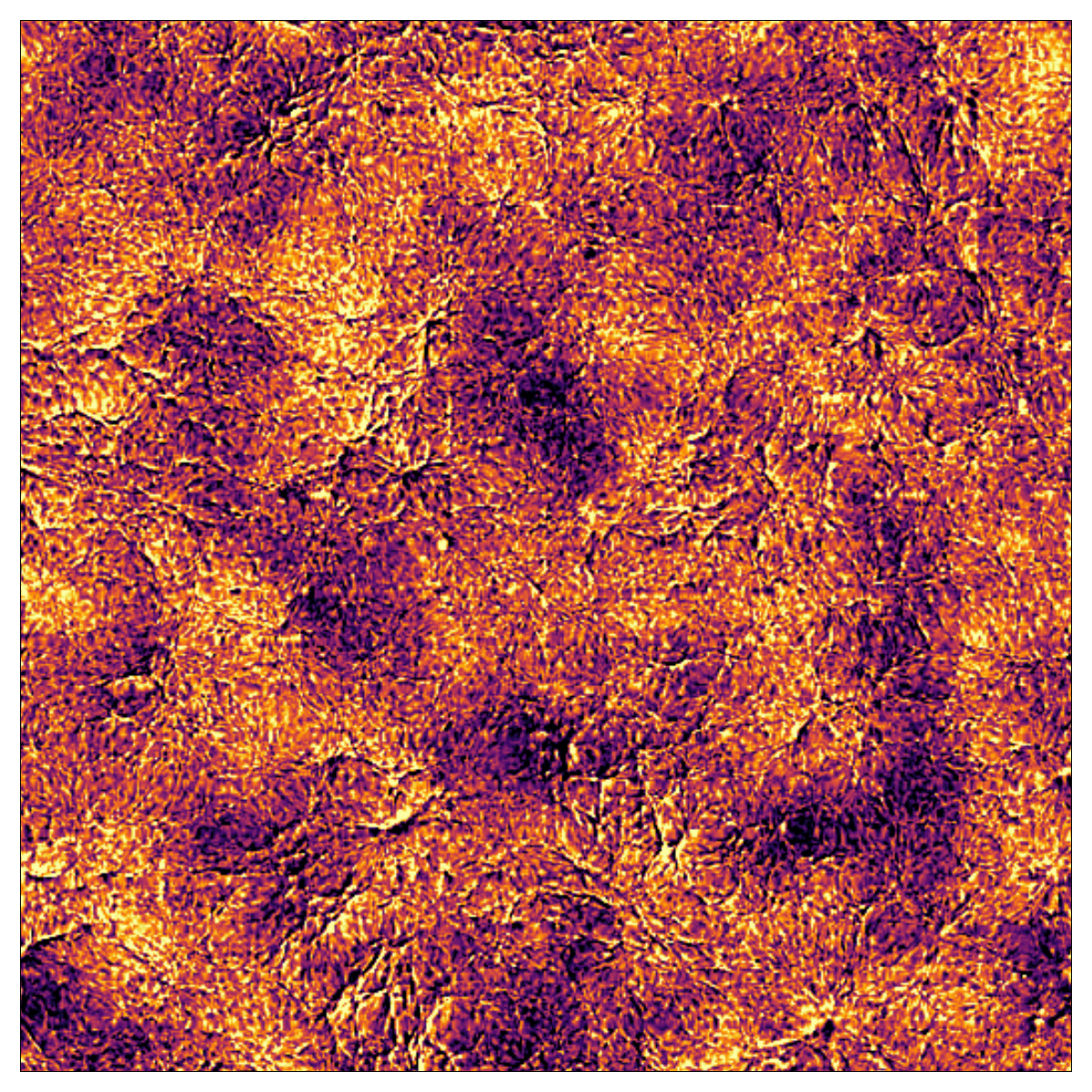}
    \end{tabular}
    \caption{Field-level visualization of BAOs in the LR and SR fields, $19\ \text{Mpc}/h$ depth 2-dimensional projections. We constructed two $410\ \text{Mpc}/h$ length LR boxes with the same initial seed, one with BAOs and one without, and we generated their respective SR fields. Shown are the LR and SR BAOs in position space, computed as the difference between fields with and without BAOs. Large-scale fluctuations are emulated from the LR conditional to the SR model output, even though the diffusion model only trained on $37\ \text{Mpc}/h$ length boxes.}
    \label{fig:BAO_dif}
\end{figure}

\begin{figure}
    \centering
    \hspace{0.9cm}Power spectra with and without BAOs
    
    \includegraphics[width=0.495\textwidth]{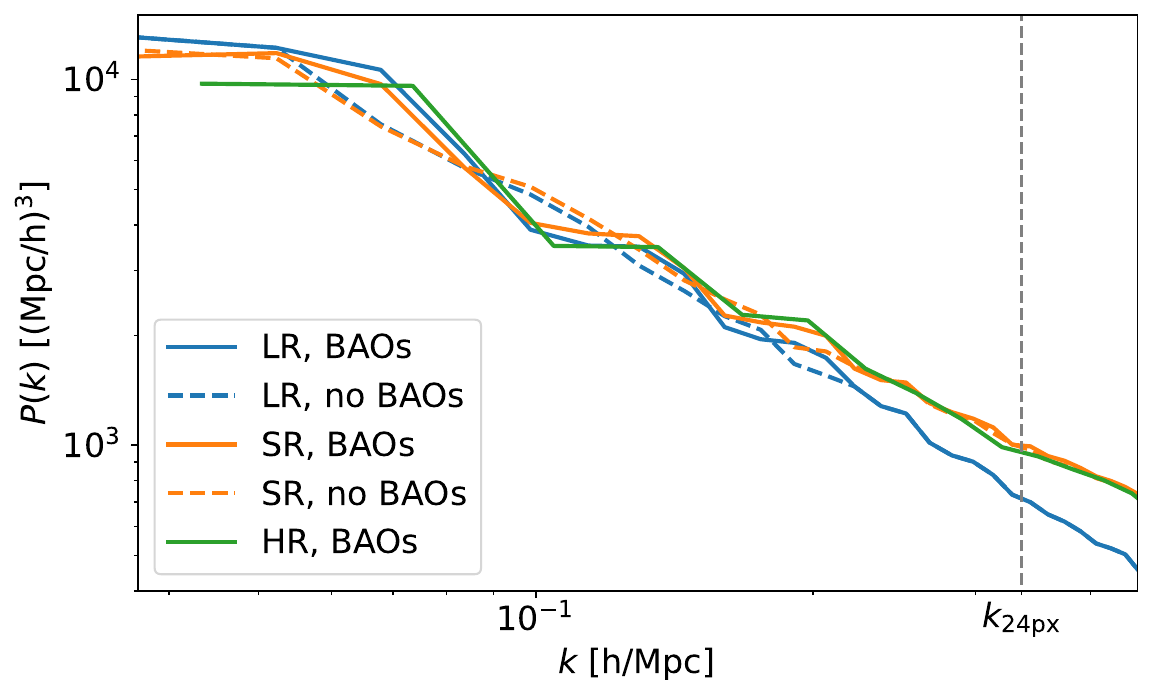}
    \caption{Power spectra for fields with and without BAOs, plotted as solid and dashed lines, respectively. When the LR conditional contains BAOs, they are emulated to SR, matching the HR training data. Conversely, an LR conditional without BAOs generates an SR field without BAOs. The SR model output emulates structure in the LR conditional at Fourier modes smaller than the $k_{24\text{px}}$ outpainting scale.}
    \label{fig:BAO_dif_ps}
\end{figure}

We further illustrate this successful large-scale response of our model by comparing fields with and without BAOs. We constructed two $410\ \text{Mpc}/h$ length LR boxes with the same initial seed, one with BAOs and one without. We have then generated their respective SR fields, and the one containing BAOs was presented in Sec~\ref{sec:results}. We show field-level images of BAOs in Fig.~\ref{fig:BAO_dif}, computed as the difference between the BAO and non-BAO fields. We see that large-scale fluctuations contained in the LR conditional field are emulated to the SR model output. (The SR fields additionally have high-frequency phase differences irrelevant to the BAOs, explained in Sec.~\ref{sec:results_variety}.) In Fig.~\ref{fig:BAO_dif_ps}, we plot power spectra of our LR and SR fields comparing the existence of BAOs. The SR power spectra follow the same pattern as the LR power spectra below $k_{24\text{px}}$, whether there are BAOs or not. Large length structures are not learned from the HR field with our outpainting method, and must accurately represented in the LR field in order to be generated in the SR field.

\section{Varying cosmologies in the LR conditional}
\label{sec:variable_cosmos}

In this Appendix, we generate new SR simulations with varied cosmological parameters in the LR conditional field as to examine the model's robustness to out-of-distribution conditional data. We vary each of $\sigma_8$ and $\Omega_\text{M}$ by $\pm5\%$ in the LR conditional from their fiducial values of $\sigma_8=0.8159$, $\Omega_\text{M}=0.3089$, giving four experiments to test the sensitivity to out-of-distribution data. For our four varied cosmologies along with a fiducial cosmology, we simulate $205\ \text{Mpc}/h$ length boxes of $128^3$ dark matter particles to build the LR conditional, each with the same initial seed. 

For each varied cosmology, we independently generate 8 SR fields, each of length $112\ \text{Mpc}/h$ ($144$~px) consisting of $5^3$ outpainting iterations. The 8 fields are generated from different volumes of the $205\ \text{Mpc}/h$ conditional. By generating several SR fields, rather than a single large field, we get an estimate of the variance of our diffusion model output.

\begin{figure}
    \centering
    \begin{tabular}{cc}
        \hspace{.6cm}$\sigma_8^-=0.7751$ & \hspace{.6cm}$\sigma_8^+=0.8567$\\
        \includegraphics[width=0.498\textwidth]{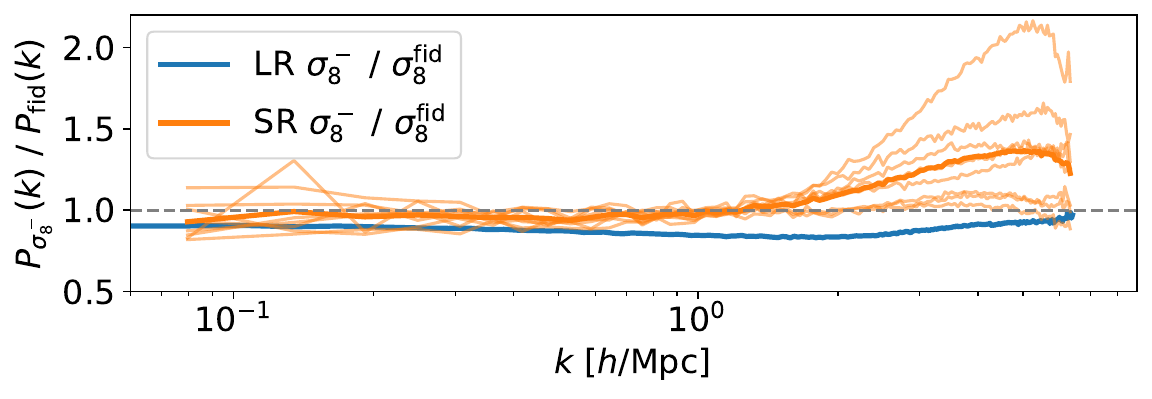} & \includegraphics[width=0.498\textwidth]{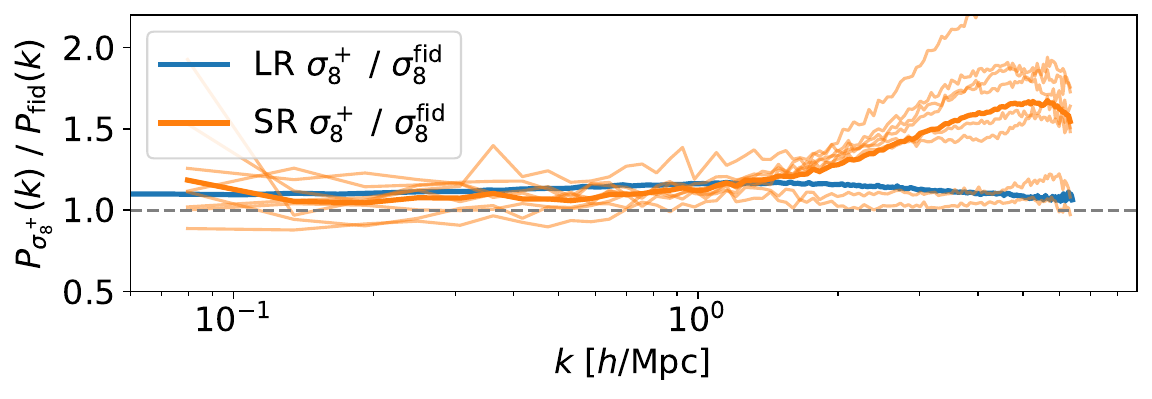}\\
        \hspace{.6cm}$\Omega_\text{M}^-=0.2935$ & \hspace{.6cm}$\Omega_\text{M}^+=0.3243$\\
        \includegraphics[width=0.498\textwidth]{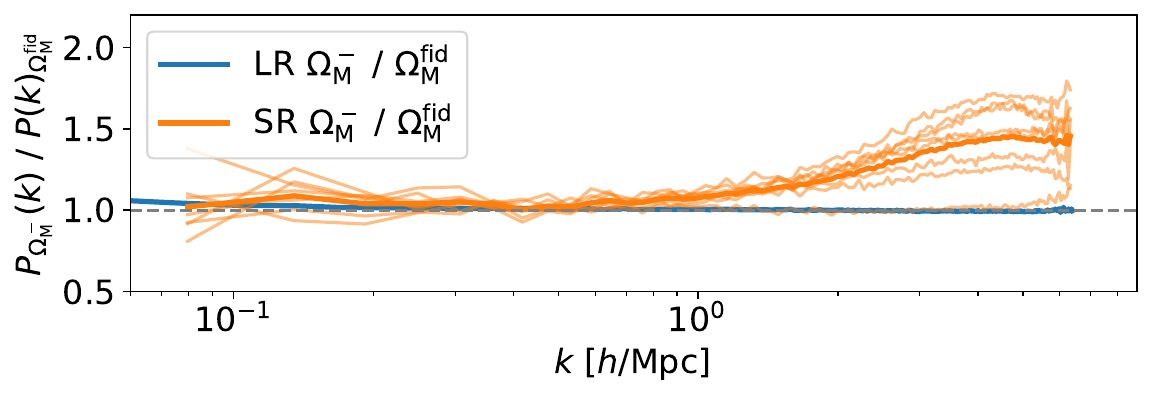} & \includegraphics[width=0.498\textwidth]{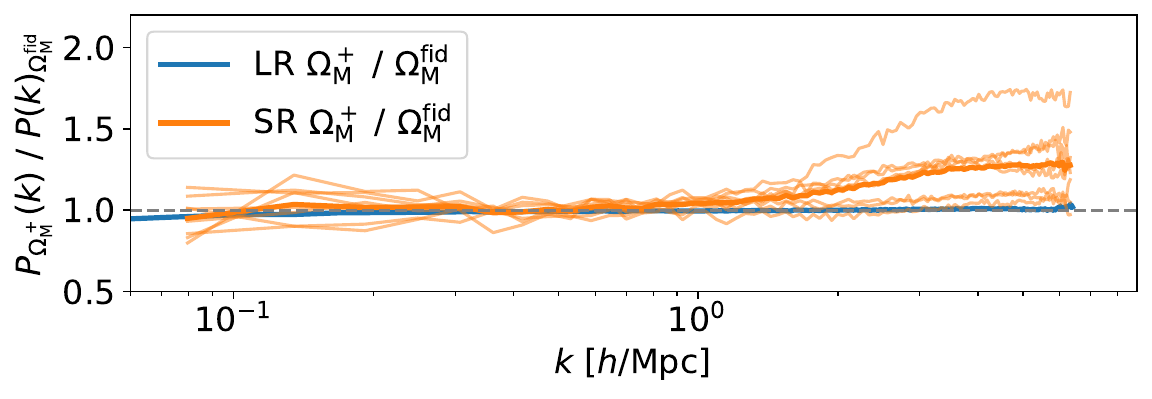}\\
    \end{tabular}
    \caption{Power spectra for varying $\sigma_8$ and $\Omega_\text{M}$ by $\pm5\%$ in the LR conditional field, shown as ratios over the fiducial power spectra with $\sigma_8=0.8159$, $\Omega_\text{M}=0.3089$, for 8 independently generated volumes, the mean denoted by the bold line. In all four varied cosmologies, the SR field tends to have increased power at nonlinear scales. The model seems to be highly sensitive to out-of-distribution conditional fields.}
    \label{fig:ps_ratios}
\end{figure}

Here we use $T=2000$ denoising steps (increased from $1250$ previously) and use a noise schedule with $\beta_0=10^{-6}$, $\beta_T=10^{-2}$. Increasing the number of steps while reducing the $\beta_t$ between steps may possibly reduce errors in this out-of-distribution regime, but we did experiment with either noise schedule and obtained qualitatively similar results. We show power spectra ratios between the varied and fiducial SR fields for the four cosmologies in Fig.~\ref{fig:ps_ratios}, along with the LR power spectra.

For $\sigma_8^-$, we expect the power spectrum to be decreased at nonlinear scales by a factor of about $0.95^2$, but our SR power spectrum only decreases slightly at these scales. There is a bit of success for $\sigma_8^+$, as we can see that the model output has nearly the proper $1.05^2$ increase in power for $k<1\ h/\text{Mpc}$. However, a major issue is that both $\sigma_8^-$ and $\sigma_8^+$ significantly increase in power above $k=1\ h/\text{Mpc}$, but we expect them to move in opposite directions. 
Considering $\Omega_\text{M}^-$ and $\Omega_\text{M}^+$, our SR model output has remained the same at nonlinear scales in the power spectrum. However, the model output again has a significant power increase above $k=1\ h/\text{Mpc}$ for both $\Omega_\text{M}^-$ and $\Omega_\text{M}^+$.

In each of the four cosmologies, the SR model output did not accurately follow the expected power spectra curve compared to the fiducial cosmology, but rather the power at larger $k$ tends to incorrectly increase significantly. We suspect this is a signature of pixel-wise noise not being properly denoised by the diffusion model. The diffusion model seems to be highly sensitive to out-of-distribution conditional data and is not robust to emulating new cosmologies from the LR conditioner. To include cosmological parameter dependence, we will thus need training data that varies these parameters.

\section{U-net diagram}
\label{sec:unet_figure}

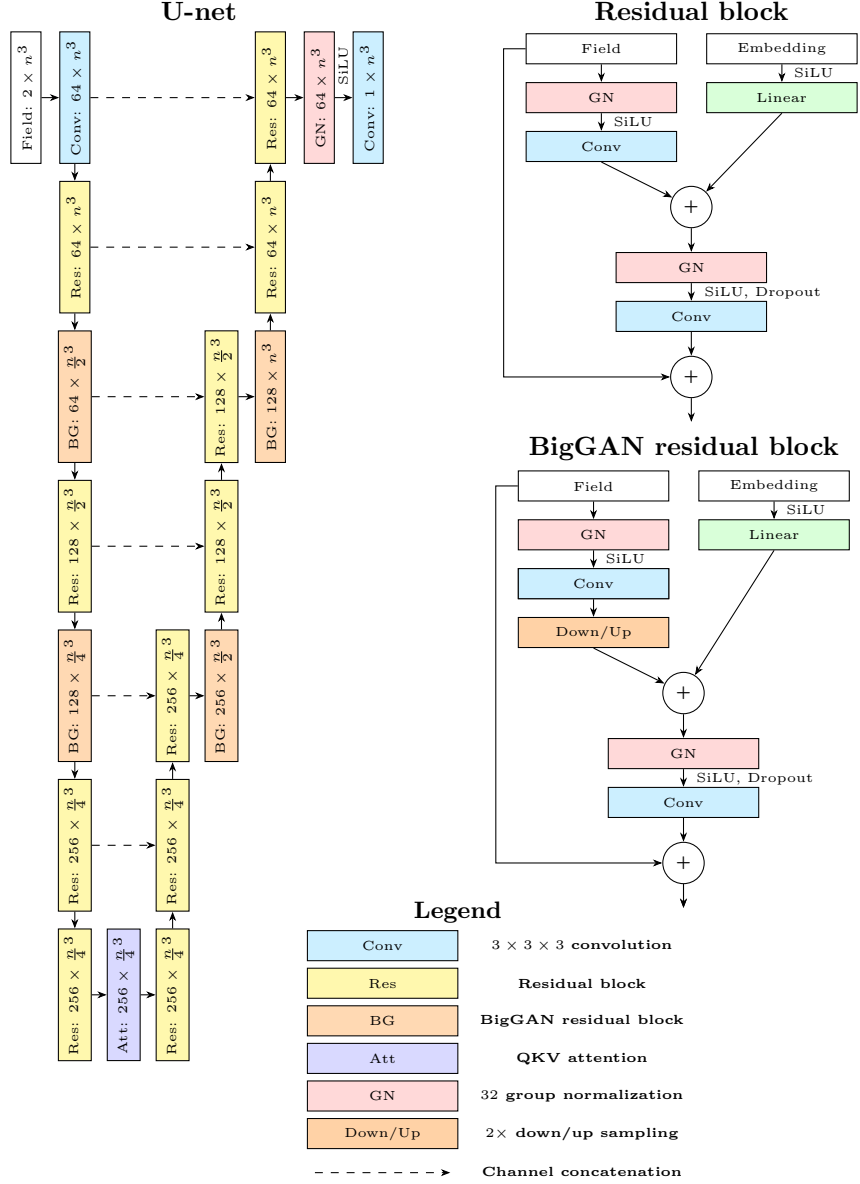
\begin{figure}[h]
\begin{center}
\begin{tikzpicture}[node distance=0.65cm]
\def\xs{5.6em}
\def\resfill{yellow!40}
\def\bgfill{orange!30}
\def\catfill{white}
\tiny

\normalsize
\node[align=center, font=\bfseries] (titleunet){U-net};
\tiny

\node (input) [cnnr, left of=titleunet, xshift=-.5cm, yshift=2.3cm] {Field: $2\times n^3$};

\node (cnn0) [cnnr, below of=input, fill=azul!75] {Conv: $64\times n^3$};
\node (resd0) [cnnr, left of=cnn0, xshift=-\xs, fill=\resfill] {Res: $64\times n^3$};
\node (resd1) [cnnr, left of=resd0, xshift=-\xs, fill=\bgfill] {BG: $64\times\frac{n}{2}^3$};
\node (resd2) [cnnr, left of=resd1, xshift=-\xs, fill=\resfill] {Res: $128\times\frac{n}{2}^3$};
\node (resd3) [cnnr, left of=resd2, xshift=-\xs, fill=\bgfill] {BG: $128\times\frac{n}{4}^3$};
\node (resd4) [cnnr, left of=resd3, xshift=-\xs, fill=\resfill] {Res: $256\times\frac{n}{4}^3$};

\node (resm0) [cnnr, left of=resd4, xshift=-\xs, fill=\resfill] {Res: $256\times\frac{n}{4}^3$};
\node (att) [cnnr, below of=resm0, fill=blue!15] {Att: $256\times\frac{n}{4}^3$};
\node (resm1) [cnnr, below of=att, fill=\resfill] {Res: $256\times\frac{n}{4}^3$};

\node (resu0) [cnnr, right of=resm1, xshift=\xs, fill=\resfill] {Res: $256\times\frac{n}{4}^3$};
\node (resu1) [cnnr, right of=resu0, xshift=\xs, fill=\resfill] {Res: $256\times\frac{n}{4}^3$};
\node (resu2) [cnnr, below of=resu1, fill=\bgfill] {BG: $256\times\frac{n}{2}^3$};
\node (resu3) [cnnr, right of=resu2, xshift=\xs, fill=\resfill] {Res: $128\times\frac{n}{2}^3$};
\node (resu4) [cnnr, right of=resu3, xshift=\xs, fill=\resfill] {Res: $128\times\frac{n}{2}^3$};
\node (resu5) [cnnr, below of=resu4, fill=\bgfill] {BG: $128\times n^3$};
\node (resu6) [cnnr, right of=resu5, xshift=\xs, fill=\resfill] {Res: $64\times n^3$};
\node (resu7) [cnnr, right of=resu6, xshift=\xs, fill=\resfill] {Res: $64\times n^3$};

\node (out0) [cnnr, below of=resu7, fill=red!15] {GN: $64\times n^3$};
\node (out1) [cnnr, below of=out0, fill=azul!75] {Conv: $1\times n^3$};

\draw [arrow] (input) to (cnn0);

\draw [arrow] (cnn0) to (resd0);
\draw [arrow] (resd0) to (resd1);
\draw [arrow] (resd1) to (resd2);
\draw [arrow] (resd2) to (resd3);
\draw [arrow] (resd3) to (resd4);
\draw [arrow] (resd4) to (resm0);

\draw [arrow] (resm0) to (att);
\draw [arrow] (att) to (resm1);

\draw [arrow] (resm1) to (resu0);
\draw [arrow] (resu0) to (resu1);
\draw [arrow] (resu1) to (resu2);
\draw [arrow] (resu2) to (resu3);
\draw [arrow] (resu3) to (resu4);
\draw [arrow] (resu4) to (resu5);
\draw [arrow] (resu5) to (resu6);
\draw [arrow] (resu6) to (resu7);

\draw [arrow, dashed] (cnn0) to (resu7);
\draw [arrow, dashed] (resd0) to (resu6);
\draw [arrow, dashed] (resd1) to (resu4);
\draw [arrow, dashed] (resd2) to (resu3);
\draw [arrow, dashed] (resd3) to (resu1);
\draw [arrow, dashed] (resd4) to (resu0);

\draw [arrow] (resu7) to (out0);
\draw [arrow] (out0) to (out1) node[xshift=-0.33cm, rotate=90] {\hspace{.85cm}SiLU};

\normalsize
\node[align=center, font=\bfseries, right of=titleunet, xshift=6cm] (titleres){\hspace{-.2cm}Residual block};
\tiny

\node (cnn0) [cnn, below of=titleres, xshift=-1.3cm, yshift=.15cm] {Field}; %: $c_\text{in}\times n^3$};
\node (cnn1) [cnn, below of=cnn0, fill=red!15] {GN}; %: $c_\text{out}\times n^3$};
\node (cnn2) [cnn, below of=cnn1, fill=azul!75] {Conv}; %: $c_\text{out}\times n^3$};

\node (emb0) [cnn, right of=cnn0, xshift=1.75cm] {Embedding}; %: $128$};
\node (emb1) [cnn, below of=emb0, fill=green!15] {Linear}; %: $c_\text{out}$};

\node (plus0) [circ, below of=cnn2, xshift=5em, yshift=-.65em] {$\bm{+}$};
\node (cnn3) [cnn, below of=plus0, yshift=-.65em, fill=red!15] {GN}; %: $c_\text{out}\times n^3$};
\node (cnn4) [cnn, below of=cnn3, fill=azul!75] {Conv}; %: $c_\text{out}\times n^3$};
\node (plus1) [circ, below of=cnn4, yshift=-.65em] {$\bm{+}$};
\node (cnn5) [cnn, below of=plus1, yshift=-.65em, draw=none] {};

\draw [arrow] (cnn0) to (cnn1);
\draw [arrow] (cnn1) to (cnn2) node[yshift=0.33cm] {\hspace{.85cm}SiLU};

\draw [arrow] (emb0) to (emb1) node[yshift=0.33cm] {\hspace{.85cm}SiLU};

\draw [arrow] (cnn2.south) to (plus0);
\draw [arrow] (emb1.south) to (plus0);

\draw [arrow] (plus0) to (cnn3);
\draw [arrow] (cnn3) to (cnn4) node[yshift=0.31cm] {\hspace{1.9cm}SiLU, Dropout};
\draw [arrow] (cnn4) to (plus1);

\coordinate (topleft) at ([xshift=-1.3cm]cnn0);
\coordinate (bottomleft) at ([xshift=-2.495cm]plus1);
\draw [arrow] (cnn0) to (topleft) to (bottomleft) to (plus1);

\draw [arrow] (plus1) to (cnn5);

\normalsize
\node[align=center, font=\bfseries, below of=plus1, yshift=-.3cm] (titlebg){\hspace{-.2cm}BigGAN residual block};
\tiny

\node (cnn0) [cnn, below of=titlebg, xshift=-1.3cm, yshift=.15cm] {Field};
\node (cnn1) [cnn, below of=cnn0, fill=red!15] {GN};
\node (cnn2) [cnn, below of=cnn1, fill=azul!75] {Conv};
\node (cnn2) [cnn, below of=cnn1, fill=azul!75] {Conv};
\node (pool) [cnn, below of=cnn2, fill=orange!35] {Down/Up};

\node (emb0) [cnn, right of=cnn0, xshift=1.75cm] {Embedding};
\node (emb1) [cnn, below of=emb0, fill=green!15] {Linear};

\node (plus0) [circ, below of=pool, xshift=5em, yshift=-.65em] {$\bm{+}$};
\node (cnn3) [cnn, below of=plus0, yshift=-.65em, fill=red!15] {GN};
\node (cnn4) [cnn, below of=cnn3, fill=azul!75] {Conv};
\node (plus1) [circ, below of=cnn4, yshift=-.65em] {$\bm{+}$};
\node (cnn5) [cnn, below of=plus1, yshift=-.65em, draw=none] {};

\draw [arrow] (cnn0) to (cnn1);
\draw [arrow] (cnn1) to (cnn2) node[yshift=0.33cm] {\hspace{.85cm}SiLU};
\draw [arrow] (cnn2) to (pool);

\draw [arrow] (emb0) to (emb1) node[yshift=0.33cm] {\hspace{.85cm}SiLU};

\draw [arrow] (pool.south) to (plus0);
\draw [arrow] (emb1.south) to (plus0);

\draw [arrow] (plus0) to (cnn3);
\draw [arrow] (cnn3) to (cnn4) node[yshift=0.31cm] {\hspace{1.9cm}SiLU, Dropout};   
\draw [arrow] (cnn4) to (plus1);

\coordinate (topleft) at ([xshift=-1.3cm]cnn0);
\coordinate (bottomleft) at ([xshift=-2.495cm]plus1);
\draw [arrow] (cnn0) to (topleft) to (bottomleft) to (plus1);

\draw [arrow] (plus1) to (cnn5);

\small
\node[align=center, font=\bfseries, below of=plus1, xshift=-3cm, yshift=0cm] (titleleg){Legend};
\tiny

\node (convleg) [cnn, below of=titleleg, xshift=-1cm, yshift=.2cm, fill=azul!75] {Conv};
\node[align=left, font=\bfseries, right of=convleg, xshift=2cm] {$3\times3\times3$ convolution};
\node (resleg) [cnn, below of=convleg, yshift=.15cm, fill=\resfill] {Res};
\node[align=left, font=\bfseries, right of=resleg, xshift=2cm] {Residual block};
\node (bgleg) [cnn, below of=resleg, yshift=.15cm, fill=\bgfill] {BG};
\node[align=left, font=\bfseries, right of=bgleg, xshift=2cm] {BigGAN residual block};
\node (attleg) [cnn, below of=bgleg, yshift=.15cm, fill=blue!15] {Att};
\node[align=left, font=\bfseries, right of=attleg, xshift=2cm] {QKV attention};
\node (gnleg) [cnn, below of=attleg, yshift=.15cm, fill=red!15] {GN};
\node[align=left, font=\bfseries, right of=gnleg, xshift=2cm] {$32$ group normalization};
\node (duleg) [cnn, below of=gnleg, yshift=.15cm, fill=orange!35] {Down/Up};
\node (dulegdescrip) [align=left, font=\bfseries, right of=duleg, xshift=2cm] {$2\times$ down/up sampling};
\coordinate (concatleft)  at ([xshift=-.9cm, yshift=-.52cm]duleg);
\coordinate (concatright) at ([xshift= .9cm, yshift=-.52cm]duleg);
\draw [arrow, dashed] (concatleft) to (concatright);
\node[align=left, font=\bfseries, below of=dulegdescrip, yshift=.15cm] {Channel concatenation};

\draw [arrow] (plus1) to (cnn5);

\end{tikzpicture}
\caption{The U-net used in this work (left) is made of residual blocks (top right), with BigGAN residual blocks (center right) used to downsample and upsample the fields. Shown after every layer name in the U-net is the layer's output channels~$\times$~volume.}
\label{fig:unet}
\end{center}
\end{figure}

We show a diagram of the U-net used in this work in Fig.~\ref{fig:unet}. This U-net is similar to that used in~\cite{dhariwal2021diffusion}, built with a series of residual blocks, and downscaling and upscaling in the BigGAN residual blocks. To allow the U-net to be dependent on the diffusion step number $t$, each residual block has an embedding as an additional input. Each embedding goes through a linear layer and then gets added in the middle of the block, as shown.

\end{document}